\documentclass{emulateapj}

\usepackage{color}
\usepackage{morefloats}
\definecolor{red}{rgb}{1,0,0}
\defcitealias{gibson09}{G09}

\shorttitle{Broad Absorption Line Disappearance}
\shortauthors{Filiz Ak et~al.}

\usepackage{natbib}

\usepackage{apjfonts}
\begin{document}

\title{Broad Absorption Line Disappearance on Multi-Year Timescales in a 
Large Quasar Sample}

\author{N. Filiz Ak\altaffilmark{1,2,3} }
\author{W.~N. Brandt\altaffilmark{1,2}, P. B. Hall\altaffilmark{4},  D.~P. 
Schneider\altaffilmark{1,2}}

\author{S. ~F. Anderson \altaffilmark{5}, R.~R. Gibson\altaffilmark{5}, 
B.~F. Lundgren\altaffilmark{6}, A.~D. Myers \altaffilmark{7},
P. Petitjean\altaffilmark{8}, Nicholas~P. Ross \altaffilmark{9},   Yue Shen\altaffilmark{10}, 
D.~G. York \altaffilmark{11} }

\author{D. Bizyaev \altaffilmark{12}, J. Brinkmann \altaffilmark{12}, 
E. Malanushenko \altaffilmark{12}, D.~J. Oravetz \altaffilmark{12}, 
K. Pan \altaffilmark{12}, A.~E. Simmons \altaffilmark{12}, B.~A. 
Weaver \altaffilmark{13}}

\altaffiltext{1}{Department of Astronomy \&   Astrophysics, Pennsylvania 
State University,  University Park, PA, 16802, USA}
\altaffiltext{2}{Institute for Gravitation and the Cosmos, Pennsylvania 
State University, University Park, PA 16802, USA}
\altaffiltext{3}{Faculty of Sciences, Department of Astronomy and Space 
Sciences, Erciyes University, 38039 Kayseri, Turkey}
\altaffiltext{4}{Department of Physics and Astronomy, York University, 
4700 Keele St., Toronto, Ontario, M3J 1P3, Canada}
\altaffiltext{5}{Astronomy Department, University of Washington, Seattle, 
WA 98195, USA}
\altaffiltext{6}{Department of Physics, Yale University, New Haven, CT
06511, USA}
\altaffiltext{7}{Department of Physics and Astronomy, University of Wyoming, 
Laramie, WY 82071, USA}
\altaffiltext{8}{Universite Paris 6, Institut d'Astrophysique de Paris, 75014, 
Paris, France}
\altaffiltext{9}{Lawrence Berkeley National Laboratory, 1 Cyclotron Road,
Berkeley, CA 92420, USA}
\altaffiltext{10}{Harvard-Smithsonian Center for Astrophysics, 60 Garden St., 
MS-51, Cambridge, MA 02138, USA}
\altaffiltext{11}{Department of Astronomy \& Astrophysics, and Enrico Fermi 
Institute, The University of Chicago, 5640 S. Ellis Ave., Chicago, IL 60637, USA}
\altaffiltext{12}{Apache Point Observatory, P.O. Box 59, Sunspot, 
NM 88349-0059, USA}
\altaffiltext{13}{Center for Cosmology and Particle Physics, New York University, 
New York, NY 10003 USA}

\email{nfilizak@astro.psu.edu}

\begin{abstract}
We present 21 examples of C\,{\sc iv} Broad Absorption Line (BAL) 
trough disappearance in 19 quasars selected from systematic 
multi-epoch observations of 582 bright BAL quasars ($1.9<z<4.5$) by 
the Sloan Digital Sky Survey-I/II \hbox{(SDSS-I/II)} and \hbox{SDSS-III}. 
The observations span \hbox{1.1--3.9~yr} rest-frame timescales, longer 
than have been sampled in many previous BAL variability studies. On  
these timescales, $\approx 2.3$\% of C~{\sc iv} BAL troughs  disappear 
and  $\approx 3.3$\% of BAL quasars show a disappearing  trough. 
These observed frequencies suggest that  many C\,{\sc iv} BAL 
absorbers spend on average at most a century along our line of 
sight to their quasar. Ten of the 19 BAL quasars showing C\,{\sc iv} 
BAL  disappearance have apparently transformed from BAL to 
non-BAL quasars;  these are the first reported examples of such 
transformations. The BAL troughs that disappear tend to be those 
with small-to-moderate equivalent widths, relatively shallow depths, 
and high outflow velocities. Other non-disappearing C\,{\sc iv} BALs in those 
nine objects having multiple troughs tend  to weaken when one of them 
disappears, indicating a connection between the disappearing and 
non-disappearing troughs, even for velocity separations as large as 
\hbox{10000--15000 $\mathrm{km\,s^{-1}}$}. We discuss possible origins 
of this connection including disk-wind rotation and changes in shielding gas. 
\end{abstract}


\section{Introduction} \label{intro}

Intrinsic absorption lines in quasar spectra are often produced by 
outflowing winds along the line of sight that are launched from the 
accretion disk or other structures around the central supermassive 
black hole \citep[SMBH; e.g.,][]{murray95,proga00}. Such absorption 
lines, shaped by the geometry and kinematic structure of the outflows, 
appear in quasar spectra as  broad absorption lines 
(BALs; $\Delta v$\,$\geq$\,2000\,$\mathrm{km\,s^{-1}}$), mini-BALs
(2000\,$\geq$\,$\Delta v$\,$\geq$\,500\,$\mathrm{km\,s^{-1}}$), or 
intrinsic narrow absorption lines 
(NALs; $\Delta v$\,$\leq$\,500\,$\mathrm{km\,s^{-1}}$); e.g., see
\citet{hamann08} and \citet{gb08}. Traditionally defined BAL troughs are 
observed in $\approx 15$\% of quasars; these troughs are sufficiently strong 
that the depths of the features lie at least 10\% under the continuum in the, 
e.g., Si\,{\sc iv}~$\lambda$\,1400, C\,{\sc iv}~$\lambda$\,1549, 
Al\,{\sc iii}~$\lambda$\,1857, or Mg\,{\sc ii}~$\lambda$\,2799 transitions 
with blueshifted velocities up to $\approx 0.1c$ (e.g., \citealp{wey91}; 
\citealp{gibson09}, hereafter G09; \citealp{allen11}; and references therein). 
Quasars that present BAL troughs in their spectra are often classified into 
three subtypes depending on the presence of absorption lines in specified 
transitions: (1) High-ionization BAL quasars show absorption lines in 
high-ionization transitions including Si\,{\sc iv}, and C\,{\sc iv}. 
(2) Low-ionization BAL quasars possess Mg\,{\sc ii} and/or Al\,{\sc iii} 
absorption lines, in addition to the high-ionization transitions. 
(3) Iron low-ionization BAL quasars show additional absorption from 
excited states of Fe\,{\sc ii} and Fe\,{\sc iii} \citep[e.g.,][and references therein]{hall02}. 

The winds revealed by intrinsic quasar absorption lines are of importance
for two main reasons. First, the observed frequency of quasar
absorption lines indicates that these winds are a significant component
of the nuclear environment. Indeed, disk accretion onto the SMBH may
require significant mass ejection for expulsion of angular momentum
from the system \citep[e.g.,][]{blandford82,crenshaw03}.
Second, quasar winds may be agents of feedback into massive galaxies,
regulating star formation and further SMBH accretion via the removal
of cold gas \citep[e.g.,][]{springel05,king10}.

BAL troughs, the strongest observed signatures of quasar winds,
often vary in equivalent width (EW) and/or shape over rest-frame
timescales of months to years
\citep[e.g.,][]{barlow93,lundgren07,gibson08,gibson10,cap11,cap12}.
Recent statistical studies of BAL variability have shown that the
fractional EW change increases with rest-frame timescale over the
range \hbox{0.05--5~yr}. Such variations could, in principle, be
driven by changes in covering factor, velocity structure, or
ionization level. Of these possibilities, the generally favored
driver for most BAL variations is changes in the covering factor
of outflow stream lines that partially block the continuum emission
\citep[e.g.,][]{hamann98, arav99}. These covering-factor changes
could ultimately be caused by rotation of a non-axisymmetric outflow
that is loosely anchored to the accretion disk, or they could arise
from large-scale changes in wind structure \citep[e.g.,][]{proga00}.
Changes in ionization level are generally disfavored as a primary
driver, since BAL troughs are often highly saturated and thus should
be only weakly responsive to ionization-level changes. Furthermore,
BAL trough variations generally do not appear to be correlated with
variations of the observable continuum (typically) longward of
Ly$\alpha$ \citep[e.g.,][]{gibson08}, although the observable
continuum is not the ionizing continuum for the BAL gas
(i.e., it is possible that the two may vary differently).

Strong variations in BAL EWs observed over multi-year 
timescales in the rest frame suggest that BAL disappearance and 
BAL emergence may be significant effects on such timescales  
\citep[e.g.,][]{gibson08,hall11}. However, largely owing to 
practical difficulties in observing large samples of BAL quasars 
on multi-year timescales (often corresponding to 
\hbox{10--20~yr} in the observed frame),  only a small number 
of BAL disappearance \citep{junk01,lundgren07,hall11,vivek12} and emergence
\citep{ma02,lundgren07,hamann08,leighly09,krongold10,cap12,vivek12}
events have been discovered. Specifically, considering the BAL 
disappearance phenomenon of most relevance to this 
paper, we are only aware of four reported cases. In the first, a 
Mg\,{\sc ii} BAL in the spectrum of the binary quasar 
\hbox{LBQS~0103--2753}  essentially disappeared over 
\hbox{$\leq 6.0$} rest-frame years, converting this object from a 
low-ionization BAL quasar to a high-ionization BAL quasar 
\citep{junk01}.%
In the second, the highest velocity C\,{\sc iv} 
trough in the spectrum of J075010.17+304032.3 nearly 
disappeared in \hbox{$\leq$ 0.3} rest-frame years, which is the only 
known example of C\,{\sc iv} disappearance \citep{lundgren07}.
In the third, the  Fe\,{\sc ii} troughs in the spectrum of 
FBQS~J1408+3054 disappeared over \hbox{$\leq$ 5.1} 
rest-frame years, converting this object from an iron 
low-ionization BAL quasar to a low-ionization BAL quasar
\citep{hall11}. In the fourth, a Mg\,{\sc ii} BAL trough
in the spectrum of SDSS J133356.02+001229.1 disappeared 
in \hbox{$\leq$ 3.7} rest-frame years and another  
Mg\,{\sc ii} BAL trough --- that emerged at higher velocity ---  
nearly disappeared in \hbox{$\leq$ 5.7} rest-frame years \citep{vivek12}. 
These disappearance examples did not involve transformations from 
BAL quasars to non-BAL quasar status because troughs from other ions, 
or additional C\,{\sc iv} troughs, remained.

We have compiled a sample of 19 quasars with 
disappearing BAL troughs from a well-defined, large sample of 
BAL quasars whose spectra were observed over \hbox{1.1--3.9~yr} 
in the rest frame. We aim to define the basic statistical 
characteristics of the BAL disappearance phenomenon. 
In this study, we will discuss the disappearance of BAL troughs as 
well as the transformation of BAL quasars to non-BAL quasars. 
These two phenomena can be distinct given that some BAL quasars 
have multiple troughs. A quasar that possesses more than one 
BAL trough could have only one of them disappear without 
transforming to a non-BAL quasar. 
We only consider BAL quasars without any remaining BAL troughs at a
later epoch (in any transition) to have transformed from a 
BAL quasar to a non-BAL quasar. 
The observations and the selection of the main sample used in this study 
are discussed in \S2. Identification of disappearing BAL troughs 
is explained in \S3. Statistical results and comparisons are presented 
in \S4. A summary of the main results and some future prospects 
are given in \S5.

Throughout this work we use a cosmology in which 
$H_0=70$~km~s$^{-1}$~Mpc$^{-1}$, 
$\Omega_M=0.3$, and 
$\Omega_{\Lambda}=0.7$. 
All timescales are in the rest frame unless otherwise mentioned. 

\section{Observations and Sample Selection} 

\subsection{Observations} \label{observations}

The Baryon Oscillation Spectroscopic Survey (BOSS), part of the  
Sloan Digital Sky Survey-III \citep[SDSS-III;][]{eisen11},  
is a five-year program \hbox{(2009--2014)} that is using the 2.5-m  
SDSS telescope \citep{gunn06} at Apache Point Observatory 
to obtain spectra for $\approx 1.5$ million luminous galaxies 
as well as $\approx 160,000$ quasars at 
$z>2.2$  \citep{anderson12,ross11}. These targets are selected from 
$\approx 10,000$~deg$^2$ of sky at high Galactic latitude. While 
the primary goal of BOSS is to obtain  precision measurements of 
the cosmic distance scale using baryon  acoustic oscillations, the 
resulting spectra (covering  \hbox{3600--10000}~\AA\ at a resolution 
of \hbox{1300--3000}; see \citealp{eisen11}) are  valuable for 
a wide range of investigations, including studies  of quasar physics.   

In addition to the main BOSS galaxy and quasar surveys, a number of 
smaller ancillary BOSS projects are being executed. One of these 
ancillary projects, relevant to this paper, focuses on studying BAL variability 
on  multi-year timescales in the rest frame. The main goal of this project 
is to move from small-sample and single-object studies of multi-year 
BAL variability to setting large-sample statistical constraints that 
can ultimately be compared with models of quasar winds. The basic 
approach is to obtain BOSS spectroscopy of BAL quasars already 
observed from \hbox{2000--2008} by \hbox{SDSS-I/II}. Most of the targets 
for this project are drawn from the \citetalias{gibson09} catalog of 5039 BAL 
quasars in the SDSS Data Release (DR)~5 quasar catalog \citep{schneider07}. 
Specifically, we have selected the 2005 BAL quasars from this catalog that  
are optically bright ($i<19.3$); have high signal-to-noise ratio  \hbox{SDSS-I/II} 
spectra (SN$_{1700}>6$ as defined by \citetalias{gibson09}, when 
SN$_{1700}$ measurements are available); have full spectral coverage 
of their C~{\sc iv},  Si~{\sc iv}, Mg~{\sc ii}, or Al~{\sc iii} BAL regions 
(implemented via the redshift cuts described in \S4 of G09); and  have 
moderately strong to strong BAL troughs (balnicity indices of 
\hbox{BI$_0>100$~km~s$^{-1}$} as defined by G09). This uniformly 
defined sample is $\approx 100$ times larger than current data sets that 
have been used to study BAL variability  on multi-year timescales in the 
rest-frame \citep[e.g.,][]{gibson08, gibson10, cap11,cap12}. We are also targeting 
102 additional BAL quasars selected to include  unusual BAL quasars 
\citep[e.g.,][]{hall02};  objects with multiple \hbox{SDSS-I/II} observations; 
and  objects with historical coverage by the Large Bright Quasar Survey 
\citep[e.g.,][]{hewett95} or the First Bright Quasar Survey 
\citep[e.g.,][]{white00}. 

Observations for our ancillary project began at the same time as the 
primary BOSS observations, and in this paper we will utilize spectra 
taken after the end of hardware commissioning 
\citep[MJD 55176; see][]{ross11} until MJD 55811 (i.e., 2009 December 
11 until 2011 September 7). Within this date 
range, 692 of our 2005 primary BAL quasars were targeted, 
and observations for this program will continue throughout the BOSS 
project. 

\subsection{Sample Selection} \label{samples}

For the purpose of the BAL disappearance studies in this paper, 
we have selected a subset of the 2005 targeted BAL quasars with 
both \hbox{SDSS-I/II} and BOSS observations that satisfy our 
selection criteria given in \S\ref{observations} --- all 102 unusual 
and other BAL quasars are excluded. We will focus on C\,{\sc iv}  
BALs, as this is the most commonly studied strong BAL transition 
and has limited blending and confusion with other nearby transitions.
Following \S4 of \citetalias{gibson09}, we therefore consider only 
those BAL quasars in the redshift range \hbox{$z=1.68$--4.93} 
where the SDSS-I/II spectra provide full coverage of the C\,{\sc iv} 
region. 

As in past work \citep[e.g.,][]{lundgren07}, we only consider C\,{\sc iv} 
BAL troughs that are significantly detached from the C\,{\sc iv} emission 
line; this selection minimizes the  chance of confusion between BAL variability 
and emission-line variability. To implement this requirement formally, we 
consider only those C\,{\sc iv} BAL trough regions lying in the velocity range 
of $-30000 \leq v_{\rm max}< -3000$~km~s$^{-1}$ ($v_{\rm max}$ is the 
maximum observed velocity for the BAL trough region and is defined fully 
in \S\ref{measur}). We show in \S\ref{ews} that this requirement should not 
significantly  bias our statistical results. The 
\hbox{$v_{\rm max}\geq -30000$~km~s$^{-1}$} requirement minimizes 
confusion between C\,{\sc iv} BALs and the Si\,{\sc iv} emission line. 

The selections above result in 582 BAL quasars which we  define as our 
``main sample''. By construction, all of these quasars have  at least one 
\hbox{SDSS-I/II} spectrum and one BOSS spectrum.  A significant minority 
have additional \hbox{SDSS-I/II} or BOSS  spectra. In total, we have 1396 
spectra for the 582 main-sample BAL quasars. The rest-frame time intervals 
between spectral  observations range from 0.26~days to 3.86~yr, although in 
this  paper our emphasis will be on $>1$~yr timescales. 

\begin{figure}[t!]
\epsscale{1.2}
\plotone{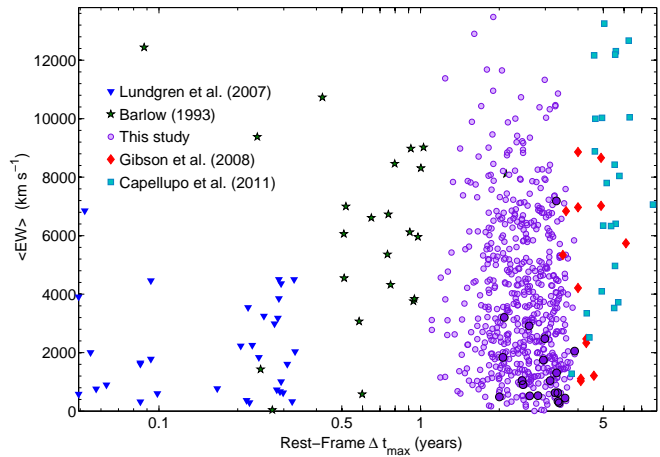}
\caption{Comparison of the total rest-frame EWs (summed over all BAL 
troughs and  averaged over two epochs) of the C\,{\sc iv} BALs for the 
sources in \citeauthor{barlow93} (1993; green stars), \citeauthor{lundgren07} 
(2007; blue triangles),  \citeauthor{gibson08} (2008; red  diamonds),  
\citeauthor{cap11} (2011; cyan squares), and our study of the 582 BAL quasars 
in the main sample (purple circles). The $x$-axis shows the maximum sampled 
timescale in the rest frame. Dark purple circles denote the quasars with 
disappearing troughs. The EW is shown in units of 
$\mathrm{km\,s^{-1}}$; 1\,$\mathrm{\AA}$ is about 
200\,$\mathrm{km\,s^{-1}}$ in the C\,{\sc iv} region. \label{fig1}}
\end{figure}

Figure \ref{fig1} shows the average total EW versus the maximum sampled 
rest-frame timescale, $\Delta t_\mathrm{max}$. The EW values of all 
C\,{\sc iv} BAL troughs in a given spectrum  are summed and averaged 
over two epochs (the first  SDSS and last BOSS spectrum). The maximum 
timescales  for our sample range between 1.10 and 3.88 yr, with a mean 
of 2.52 yr and a median of 2.50 yr. For comparison, we have also plotted 
several other samples of BAL quasars whose BAL variability properties 
have been previously investigated \citep{barlow93,lundgren07,gibson08,cap11}.
Note the relatively large size of our sample compared to past 
work as well as the fairly long timescales being probed.

We have cross correlated our 582 main-sample BAL quasars
with the \citet{shen11} catalog to check for radio emission 
detected in Very Large Array (VLA) Faint Images of the Radio 
Sky at Twenty-Centimeters \citep[FIRST;][]{becker95} observations.
\citet{shen11} list the radio-loudness parameter defined as  
$R = f_{\rm 6cm}/f_{\rm 2500{\rm \AA}}$, where  $f_{\rm 6cm}$ is 
the radio flux density at rest-frame 6 cm and 
$f_{2500{\rm \AA}}$ is the optical flux density at rest-frame 2500 
$\rm \AA$. We also have examined the NRAO VLA Sky Survey 
\citep[NVSS;][]{condon98} for three quasars that are not located 
in the FIRST footprint. Radio emission is detected from 55 of the 
main-sample quasars; seven of them are radio loud with $R \geq 100$.

\section{Identification of Disappearing BALs}

\subsection{Continuum Fit and  Normalization} \label{continuum}

To study variations in BAL characteristics, we investigate multi-epoch 
spectra that are normalized by an estimated continuum model.  To determine 
proper continuum levels, we first  corrected the main-sample  spectra for 
Galactic extinction using a Milky Way extinction model \citep{cardelli89}  
for $R_{\mathrm{V}}$ =  3.1. The $A_{\mathrm{V}}$ values were taken from 
the \citet{schlegel98}. We  then translated observed wavelengths to the rest 
frame using redshifts from \citet{hw10}.

The  data-reduction pipelines for SDSS I/II (DR7) and BOSS 
\hbox{(v5\_4\_45)}\footnote{The BOSS pipeline is described 
in \citet{DR8}.} remove night-sky lines from the data, but 
occasionally there are significant residuals near prominent 
night-sky lines.  The data-processing algorithm flags the 
afflicted pixels  in pixel-mask columns. We examine the 
``BRIGHTSKY''  mask column and remove the flagged  
pixels from each spectrum.

As in \citetalias{gibson09},  we select the following six relatively line-free 
(RLF) windows, if they have spectral coverage, to fit a continuum model 
to each spectrum: 
\hbox{1250--1350}\,\AA, \hbox{1700--1800}\,\AA, \hbox{1950--2200}\,\AA, 
\hbox{2650--2710}\,\AA, \hbox{2950--3700}\,\AA, and \hbox{3950--4050}\,\AA. 
We select these RLF windows to be free from strong emission  and absorption 
in general and  to represent the underlying continuum both blueward and redward 
of the  C\,{\sc iv} and Si\,{\sc iv} BAL regions; the use of the RLF windows spanning 
a fairly broad range of wavelengths provides useful leverage for constraining the 
continuum shape. We avoid the use of any window at shorter wavelengths due 
to heavily absorbed regions, such as H\,{\sc i} absorption in the Ly${\alpha}$ forest.  

Previous studies of BAL variability have  used a variety of models to define the 
underlying continuum, such as  power-law fits \citep[e.g.,][]{ barlow93,cap11,cap12}, 
reddened power-law fits \citep[e.g.,][]{gibson08,gibson09}, and polynomial fits 
\citep[e.g.,][]{lundgren07}. As in G09, we prefer to use a reddened power-law that 
reconstructs  the underlying continuum well at all covered wavelengths with a 
small number (three) of parameters. \citet{hopkins04} show that the intrinsic dust 
reddening in quasar spectra is dominated by Small-Magellanic-Cloud-like (SMC-like) 
reddening.
We reconstruct the continuum with a power-law model that is intrinsically reddened 
using the SMC-like reddening model from \citet{pei92}. The three continuum-model 
parameters are thus the power-law normalization, the power-law spectral index, 
and the intrinsic-absorption coefficient.  

The continuum-model parameters for each spectrum are obtained 
from a non-linear least-squares fit with an iterative ``sigma-clipping'' 
algorithm.  The sigma-clipping filtering iteratively  excludes the data 
points that deviate by more than 3$\sigma$ from the previous fit.  
Absorption or emission features (e.g., intervening absorption lines) 
that are occasionally present in the RLF windows  are thereby filtered. 
Typically 1--8\% of the RLF-window spectral pixels are excluded in a 
given spectrum. We calculate the continuum-model parameters with 
68.3\% confidence bounds  \citep[given that  the average line-flux 
sensitivity in SDSS spectra matches the $1\sigma$ noise;][]{bolton04}. 
We calculate the confidence bounds using  numerical $\Delta \chi^2$ 
confidence region estimation for three parameters of interest; see,
e.g., \S15.6.5 of \citet{press}. We propagate the derived continuum 
uncertainties through to continuum-normalized flux densities 
and  EW measurements (see \S\ref{measur} and Table 2). 

We do not assign physical meaning to the resulting continuum-model 
parameters, because   
(1)  degeneracy in model parameters allows multiple sets of parameters 
to produce nearly the same continuum, and  
(2) flux levels in blue regions (3500--5500 $\rm \AA$ in the observed 
frame) of BOSS spectra are not yet absolutely or relatively calibrated 
to better than $\pm$10\% overall, with greater uncertainties at shorter 
wavelengths owing to remaining instrumental effects 
\citep[e.g.,][]{margala11}. For further discussion of difficulties in the 
physical interpretation of continuum-model parameters, 
see \S2.1 of \citet{gibson08}.

\begin{figure}[ht!]
\epsscale{1.2}
\plotone{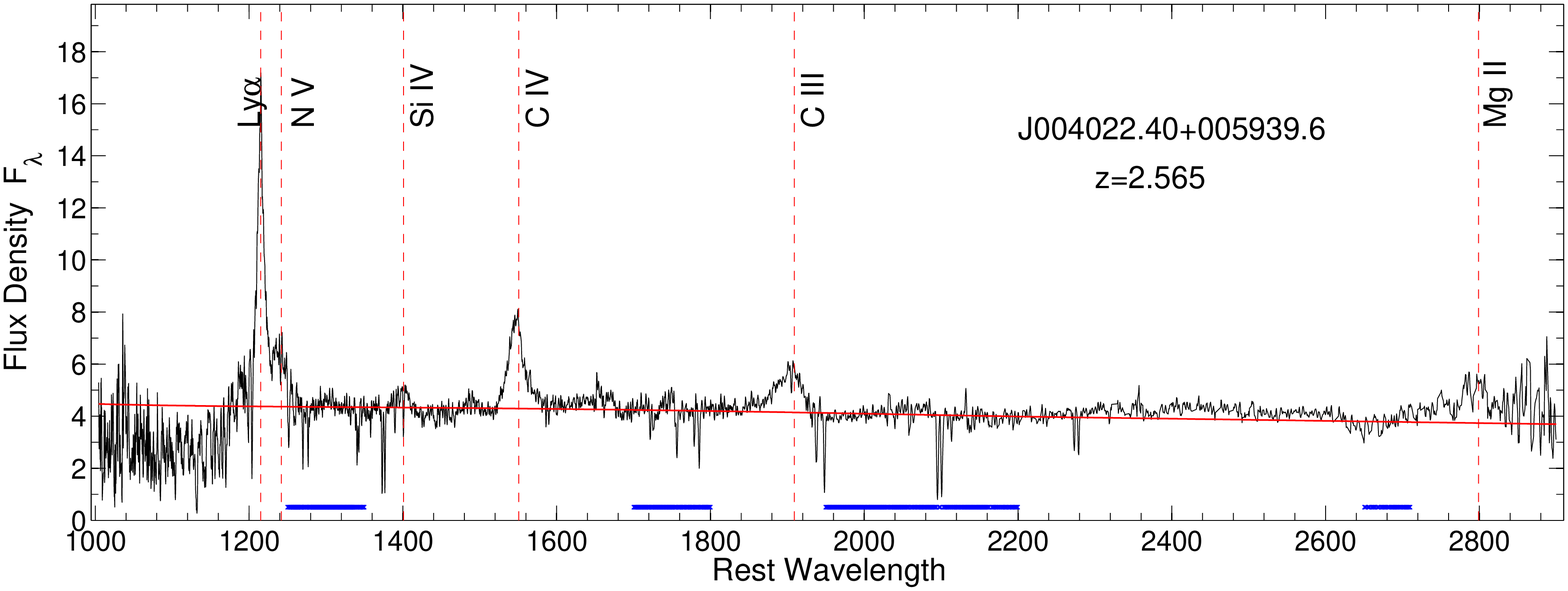}
\plotone{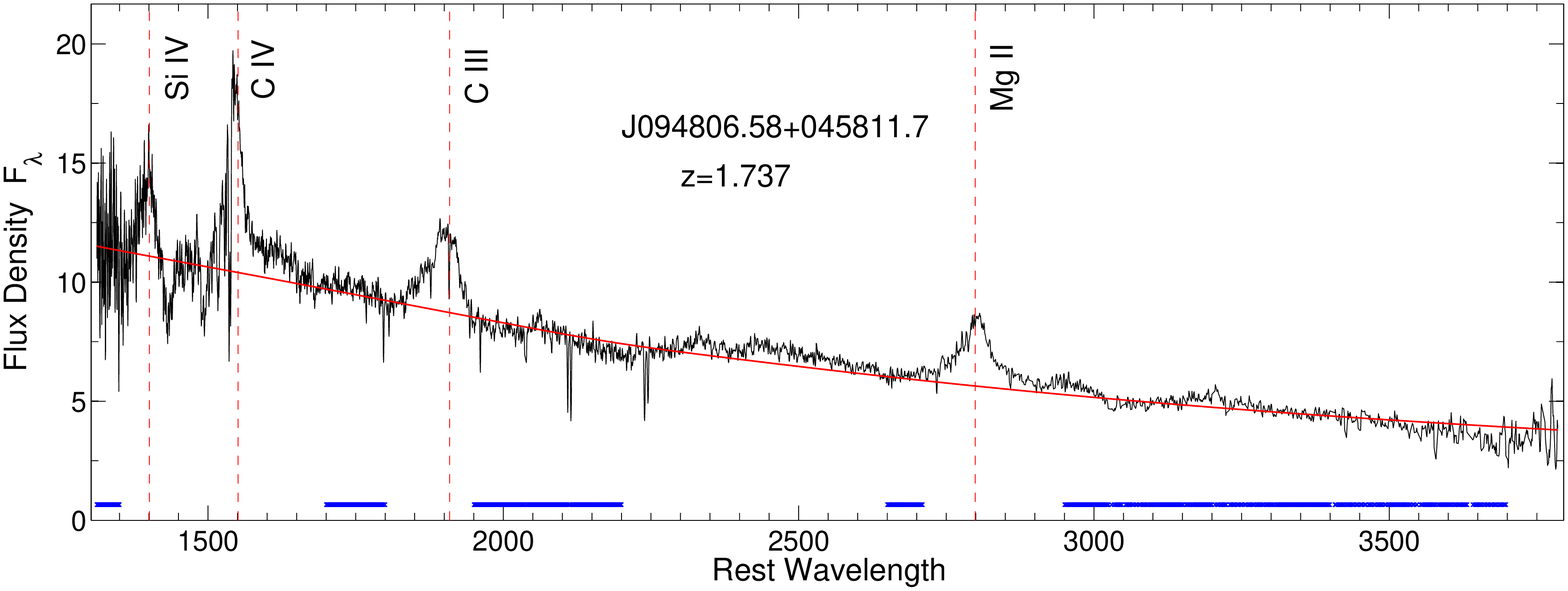}
\plotone{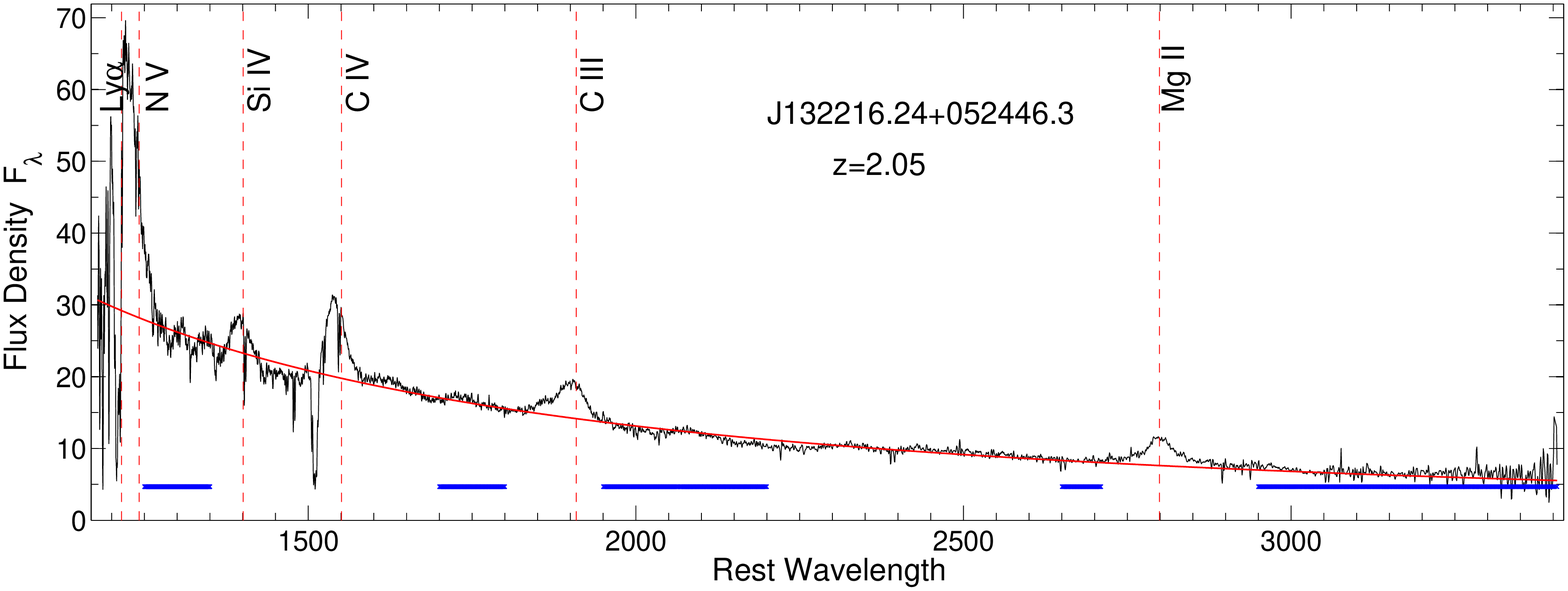}
\plotone{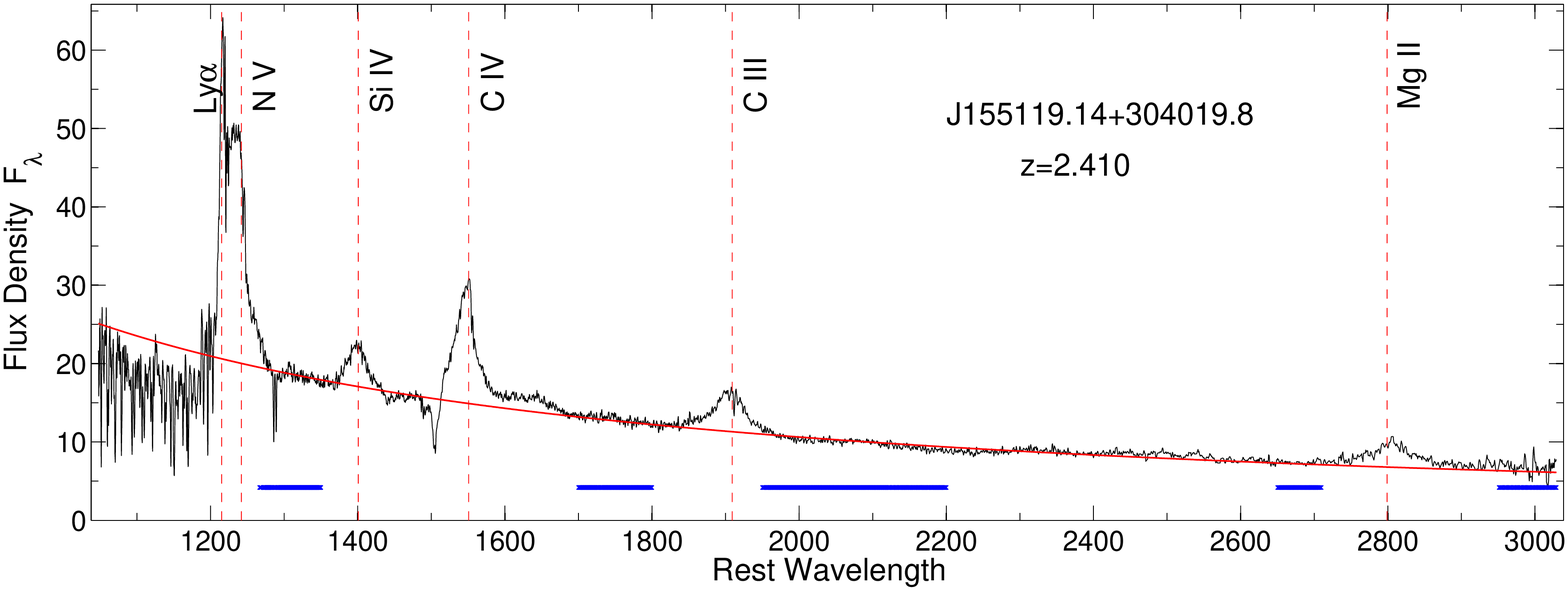}
\caption{The best-fit continuum models for four example BOSS spectra of 
quasars with disappearing BAL troughs. The observed flux density (in units 
of $\mathrm{10^{-17}~erg~cm^{-2}~s^{-1}~\AA^{-1} }$) and the rest-frame 
wavelength (in $\mathrm{\AA}$) of each object (black) is fitted with the 
continuum model (solid red) in the plotted RLF windows (blue 
horizontal lines). Discrete spectral features within the RLF 
windows were removed via the iterative continuum  fitting process. }
\label{fig2}
\end{figure}

Figure \ref{fig2} shows the best continuum-model fits for some examples of  
BOSS spectra. The RLF windows for each spectrum have miscellaneous  
line-like features that are iteratively excluded by our sigma-clipping filtering. 
Visual inspection shows that our continuum-reconstruction algorithm 
produces appropriate continuum levels. Comparison of  local continua 
blueward and redward of  C\,{\sc iv} BAL troughs (not in the RLF 
windows) shows  good agreement between the data and the fitted continuum. 
Furthermore, there is good agreement between two epochs that are fitted 
independently. The reddened power-law model 
includes extinction that changes only slowly with wavelength 
and cannot produce false BALs given these local continuum constraints. 

In addition to the statistical errors that we have quantified above,  we also 
considered possible systematic errors in the continuum fitting. For example, 
we considered the  systematic errors arising from our sigma-clipping filtering 
technique. As a test of the robustness of this approach, we repeated the 
filtering for $2\sigma$ and $4\sigma$ clipping and obtained essentially the 
same continuum fits as with $3\sigma$ clipping. We  further estimated 
systematic errors in continuum models with  one-sided sigma-clipping filtering. 
Sigma clipping of only positive deviations and of only negative deviations give  
lower and upper bounds on the continuum, respectively. We found that 
the estimated systematic errors with this particular method are less than 
the statistical errors. 
Another possible systematic error could arise when the spectral region blueward 
of the  Si\,{\sc iv} BAL region lies near the edge of the observed spectrum where 
there is significant noise. However, a visual comparison of the continuum-fitting 
results for such objects does not show any significant difference from the others.

 \subsection{Measurements of BAL Properties} \label{measur}

As is a common practice for BAL studies 
\citep[e.g.,][]{lundgren07,gibson08,gibson09}, we smoothed each 
spectrum with a Savitzky-Golay \citep[SG;][]{sg} algorithm 
before investigating the BAL troughs. The SG parameters were 
selected to perform local linear regression on three consecutive data 
points that preserve the trends of slow variations and smooth the 
fluctuations originating from noise. We used the smoothed spectra only 
to facilitate C\,{\sc iv} BAL detection. 

We searched each smoothed spectrum for absorption troughs at a 
level of $\geq$10\% under the continuum as given in the traditional 
definition of BAL and mini-BAL troughs. Velocities corresponding 
to the shortest and the longest wavelengths for a given trough, 
$v_{\rm max}$ and $v_{\rm min}$ respectively, determine each 
trough's width. We sorted the features into mini-BAL troughs 
(500--2000 $\mathrm{km\,s^{-1}}$ wide) and BAL troughs 
($\geq~$2000 $\mathrm{km\,s^{-1}}$ wide); see \S\ref{intro} for 
further discussion. 

We detected a total of 925 distinct C\,{\sc iv} BAL troughs in observations of 582 
main-sample quasars. As described in ¤2.2, we consider only troughs lying 
in the velocity range of {$-30000 \leq v_{\rm max} < -3000$~km~s$^{-1}$} 
to minimize confusion between BAL variability and emission-line variability. 
For a small fraction of the distinct C\,{\sc iv} BAL troughs (15\%), the trough 
region extends beyond the given velocity limits, mainly at the low velocity 
boundary. For such BAL troughs we assign the relevant cut-off velocity as 
the boundary of the BAL trough, and we consider the portion of the trough 
that lies within the given limits for further calculations.  A visual comparison 
of the detected BAL troughs and the positions of these troughs with those in 
G09 shows good general agreement. 

We calculate the C\,{\sc iv} balnicity index (BI$^\prime$)  of each spectrum. 
We use a similar expression to the traditional BI definition given by 
\citet{wey91}. Our expression is 
\begin{equation}
\mbox{BI$^\prime$} \equiv \int_{-3000}^{-30000} \left( {1-\frac{f(v)} {0.9}} \right) C dv
\end{equation}
In this definition, BI$^\prime$  is expressed in units of  $\mathrm{km\,s^{-1}}$ 
where $f(v)$ is normalized flux density as a function of velocity, and $C$ is a 
constant which is equal to 1.0  only where a trough is wider than 
2000\,$\mathrm{km\,s^{-1}}$, and is 0.0 otherwise. The only difference from 
the traditional balnicity definition is the limiting velocities of the C\,{\sc iv} BAL 
region (i.e., from $-$3000 to \hbox{$-$30000 $\mathrm{km\,s^{-1}}$} instead of 
from $-$3000 to \hbox{$-$25000 $\mathrm{km\,s^{-1}}$}).\footnote{Negative 
signs indicate that the BAL trough is blueshifted with respect to the systemic 
velocity.}

We calculate the rest-frame equivalent width (EW) for each BAL trough in 
units of both \AA\ and $\mathrm{km\,s^{-1}}$; note that \hbox{1\,\AA\\} 
corresponds to \hbox{$\approx $\,200\,$\mathrm{km\,s^{-1}}$} in the 
C\,{\sc iv} absorption region in the rest frame.  EW and EW uncertainties 
are calculated from unsmoothed data using Equations 1 and 2 of 
\citet{kaspi02}. In addition, following \citet{gal06}, we define the 
$f_{\rm deep}^{25}$ parameter to represent the fraction of data points 
that lie at least 25\% under the continuum level in each BAL trough. 
 
As discussed in \citet{gibson08}, BI$^\prime$ and EW measurements alone 
are not satisfactory for variability studies. For the purpose of better 
understanding spectral variations, we compared observations of a 
given quasar at two different epochs that were obtained with a time 
interval $\Delta t$. We define $\Delta t$  as the rest-frame interval 
between the  last observation showing a BAL trough (at time $t_{1}$) 
and the first observation where the trough has disappeared (at time 
$t_{2}$). Note that $\Delta t$ can be different from $\Delta t_{\rm max}$ in 
\S\ref{samples} for the quasars that are observed  in more than two 
epochs; $\Delta t$ represents an upper limit for the disappearance 
time of a BAL trough with the EW value seen in the last epoch in which 
the trough was present. For our sample, the range of $\Delta t$ is 
1.10--3.88 yr with a mean of 2.47 yr and a median of 2.45 yr.

The BOSS spectra are provided at the same observed wavelengths as the 
SDSS spectra (except for the additional BOSS wavelength coverage), therefore
we can compare spectra at identical wavelengths. Proper comparison 
requires consideration of differences in signal-to-noise ratio (S/N) 
incorporating standard deviations ($\sigma$). Therefore,  each (unsmoothed) 
pixel in the spectra, observed at $t_1$ and $t_2$,  is compared using the 
following calculation of  the quantity $N_{\sigma}$:
\begin{equation}
N_{\sigma}(\lambda) =  {\frac { f_2 - f_1 } 
{\sqrt {\sigma_2^2  + \sigma_1^2 }}}
\end{equation}
where $f_1$  and $f_2$ are the normalized flux densities  and  $\sigma_1$ 
and $\sigma_2$ are the flux density standard deviations at wavelength 
$\lambda$. Both $\sigma_1$ and $\sigma_2$ include uncertainties from the 
continuum model (see \S\ref{continuum}), in addition to the uncertainties from 
the observations. To summarize $N_{\sigma}$ is a measurement of the deviation between 
two observations for each pixel in units of $\sigma$ (see Figure~3).  

\subsection{Selection of Disappearing BALs} \label{selection}

BAL troughs are often complex and can be difficult to quantify with simple and 
standard rules. Nevertheless, we define objective criteria to search all 582 
main-sample BAL quasars for disappearing C\,{\sc iv} BAL troughs. These 
criteria were based on the comparison between two normalized spectra, 
$S_1$ and $S_2$, observed at epochs $t_1$ and $t_2$. The utilized criteria 
were the following:

\begin{itemize}

\item
The spectral region in $S_2$ corresponding to the trough seen in $S_1$ 
(lying between $v_{\rm min}$ and  $v_{\rm max}$ as defined in \S\ref{measur}, 
and corresponding to the gray shaded regions in Figure~3) should be free from 
any BALs or mini-BALs. However, we do allow the trough region in $S_2$ to 
contain residual NALs. Our upper limit for any remaining C\,{\sc iv} absorption 
in $S_2$ will be discussed below.

\item
A two-sample $\chi^2$ test \citep[see \S4.4 of][]{bevington03} comparing the 
data points in the trough region between $S_1$ and $S_2$ gives a probability 
of $P_{\chi^2}<10^{-8}$ of there being no change. This criterion establishes that 
there has been a highly significant change in the trough region between $S_1$ 
and $S_2$;  variations with less-significant values of $P_{\chi^2}$ may be real 
in some cases but can be arguable allowing for both statistical and systematic 
errors (see \S\ref{continuum}), as well as the general complexity of BAL troughs. 
The chosen probability threshold has been selected based upon visual inspection.

\end{itemize}

Following the stated criteria, we identify 21 examples of disappearing C\,{\sc iv} 
BAL troughs in 19 distinct quasars; these 19 quasars have $\Delta t$ values of 
2.0--3.3 yr. Figure~3 compares normalized SDSS and BOSS spectra in the 
C\,{\sc iv} and Si\,{\sc iv} regions for quasars with disappearing BAL troughs. 
$N{_\sigma}$ values are also plotted for each wavelength bin.

\begin{figure}[ht!]
\epsscale{1.2}
\plotone{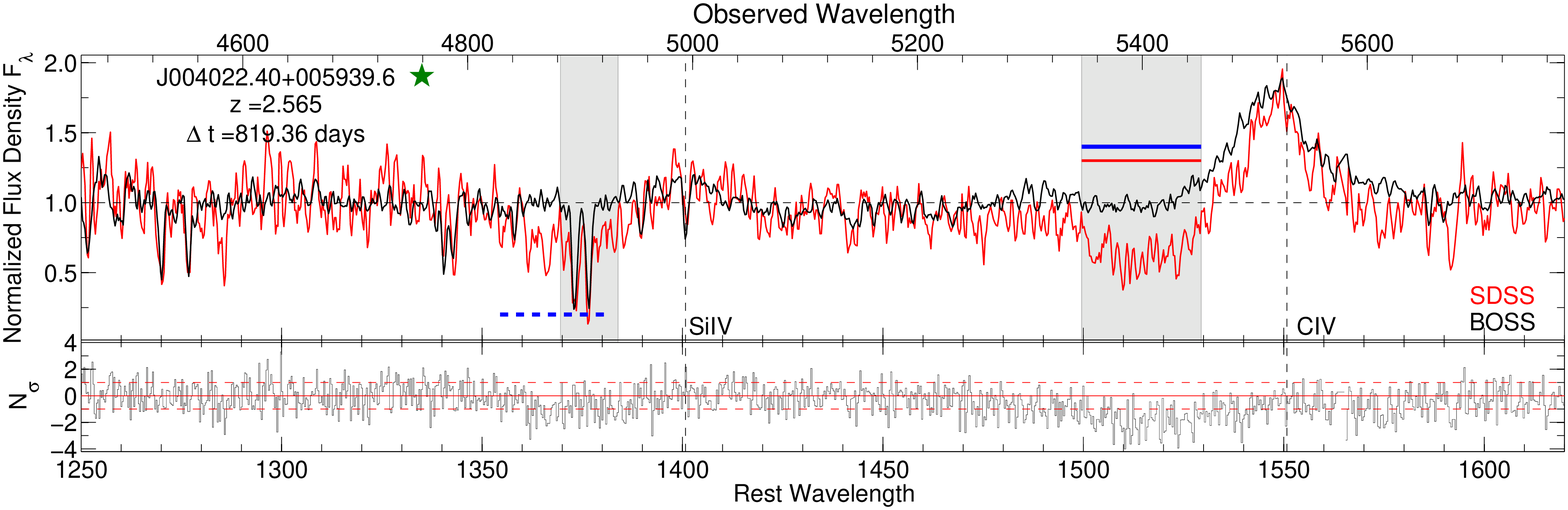}
\plotone{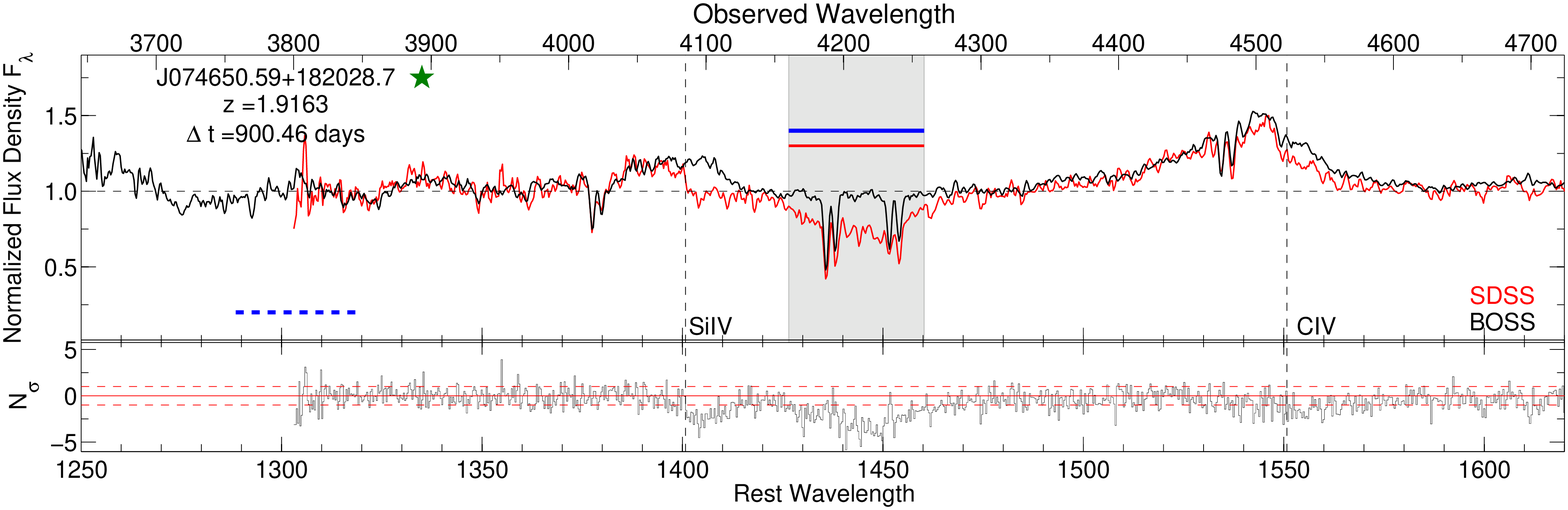}
\plotone{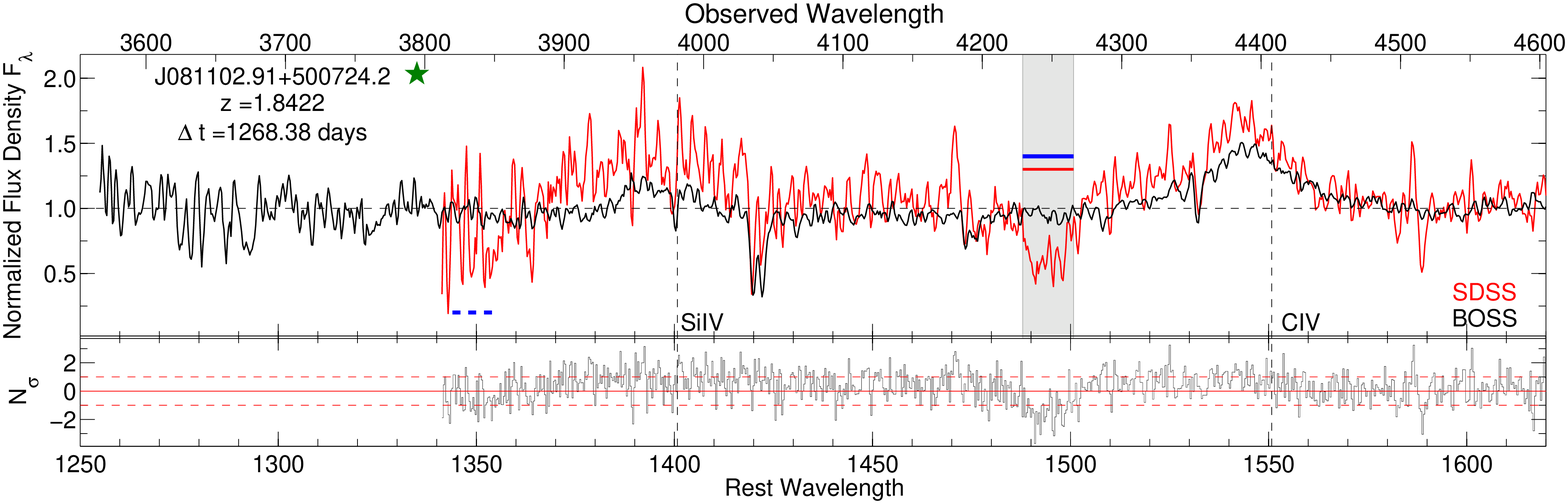}
\plotone{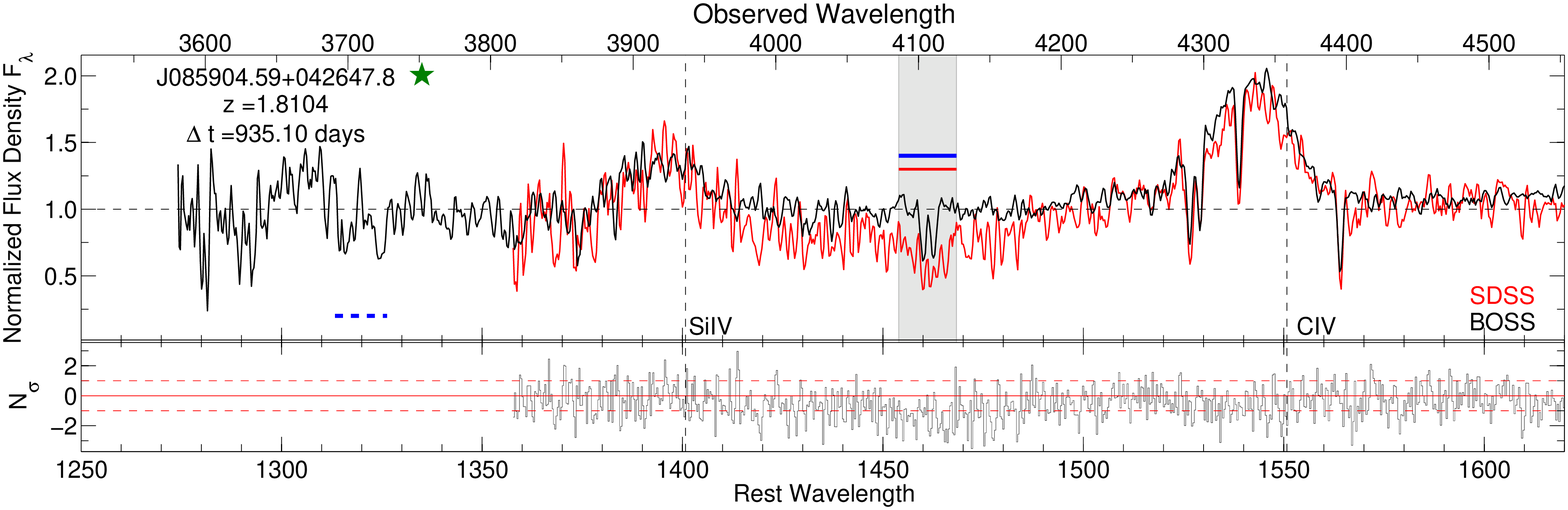}
\plotone{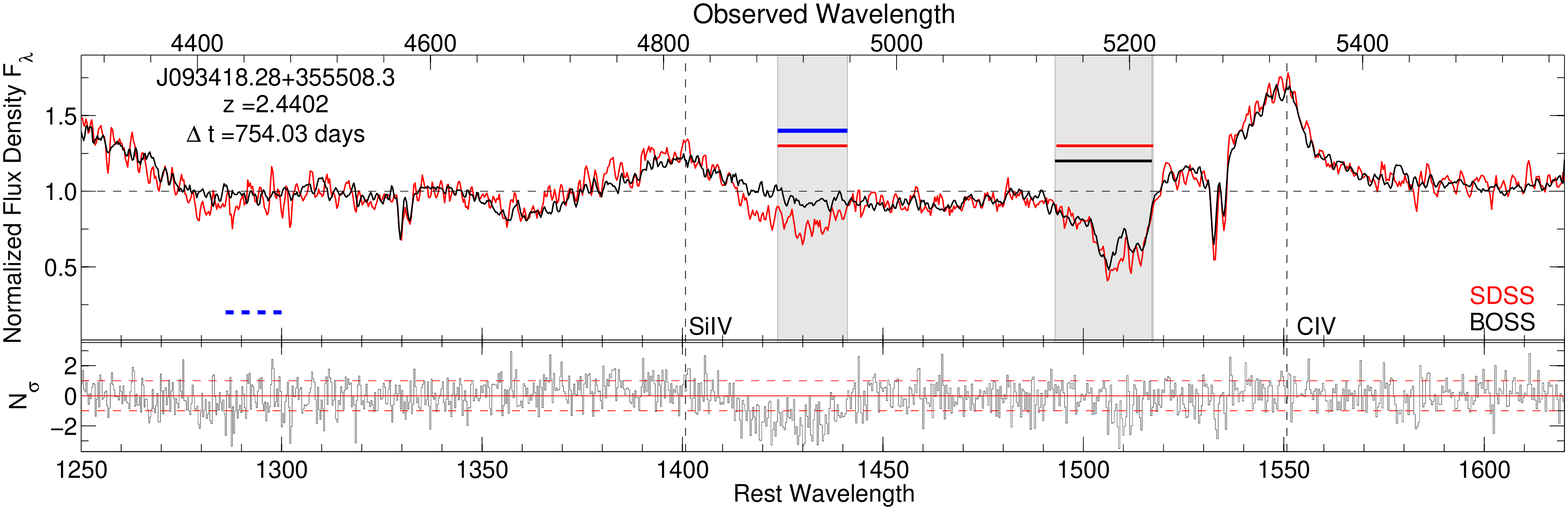}
\plotone{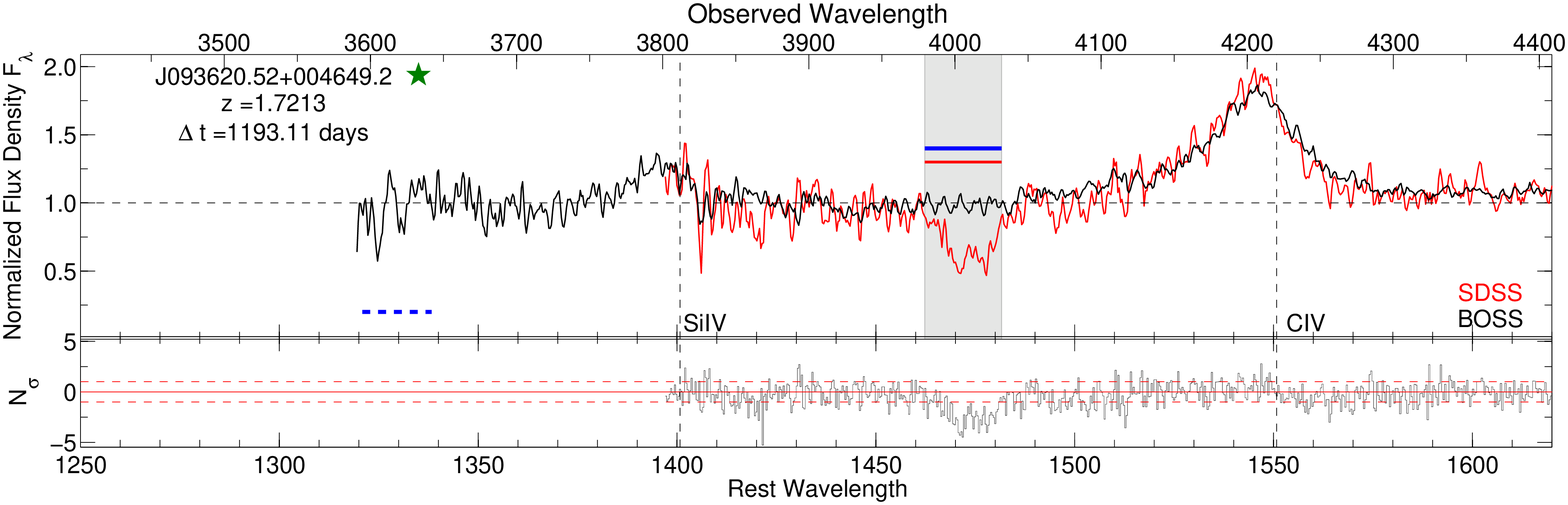}
\plotone{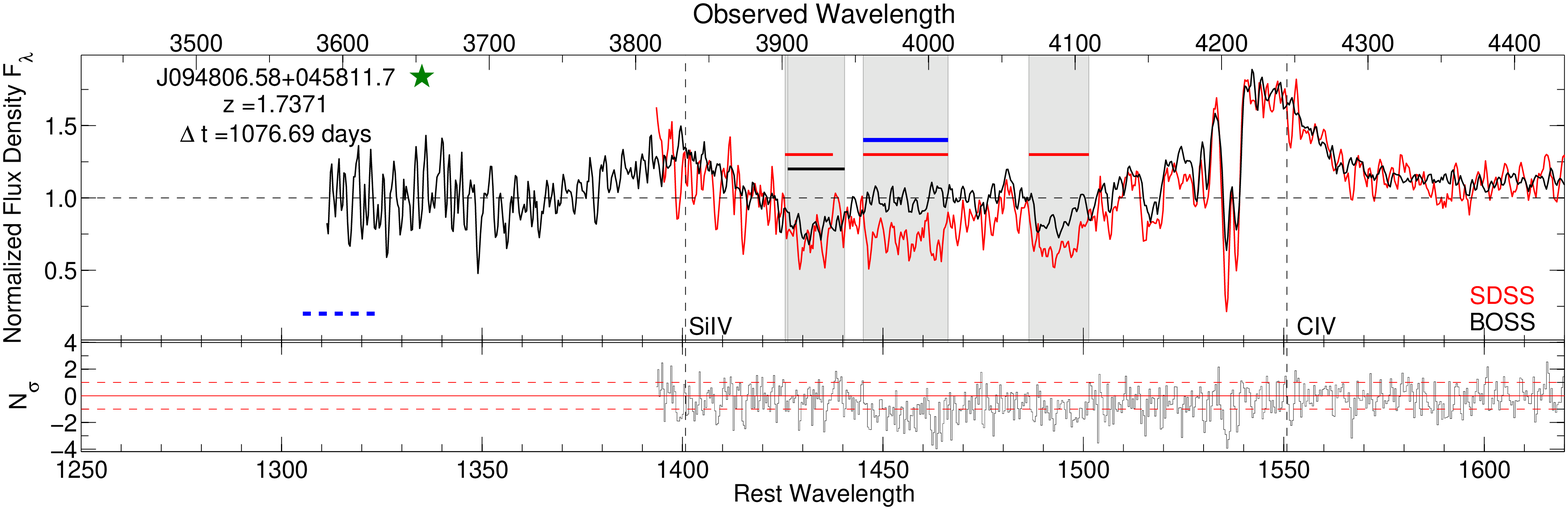}
\plotone{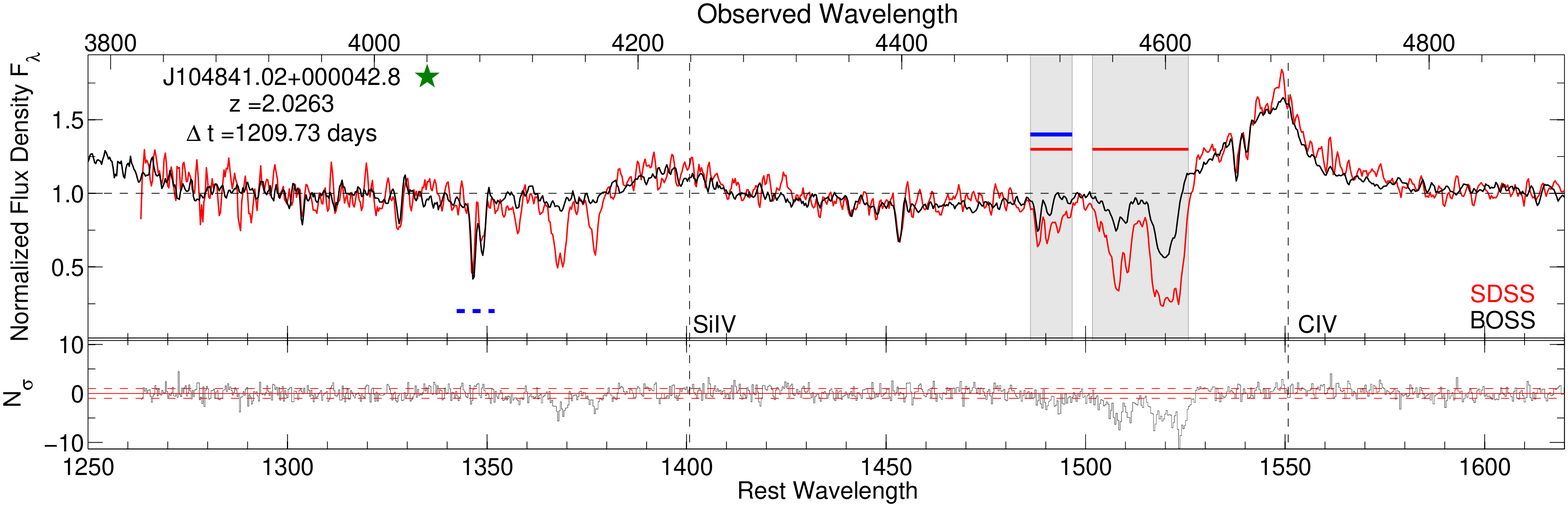}
\end{figure}

\begin{figure}
\epsscale{1.2}
\plotone{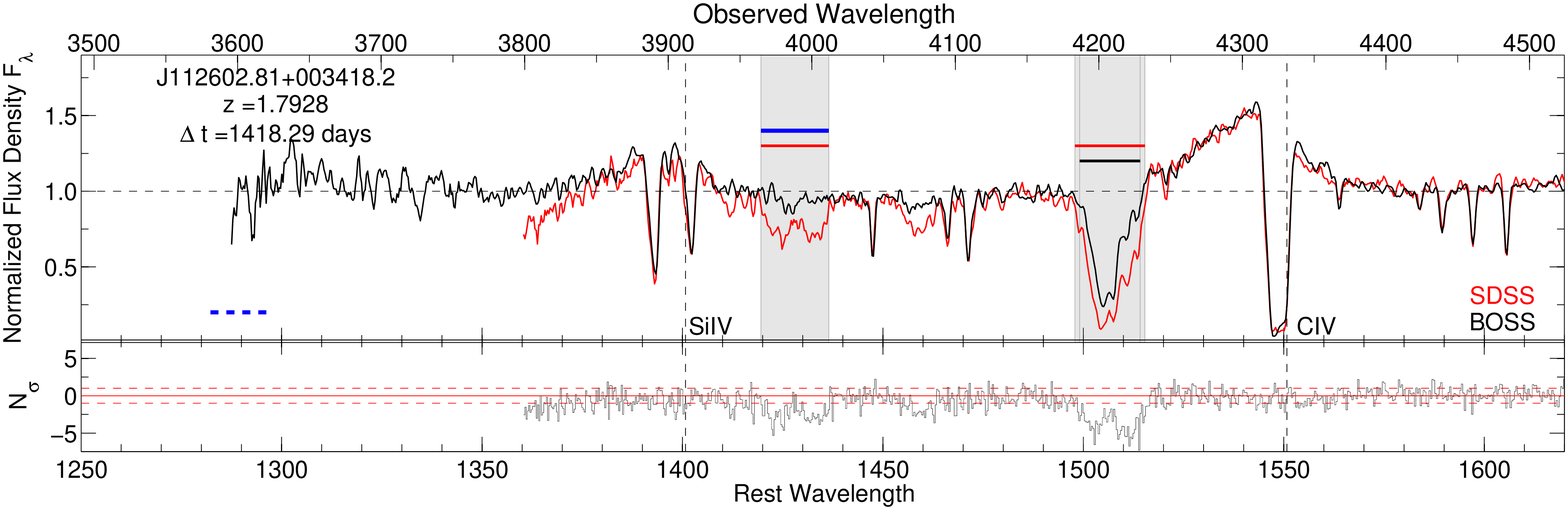}
\plotone{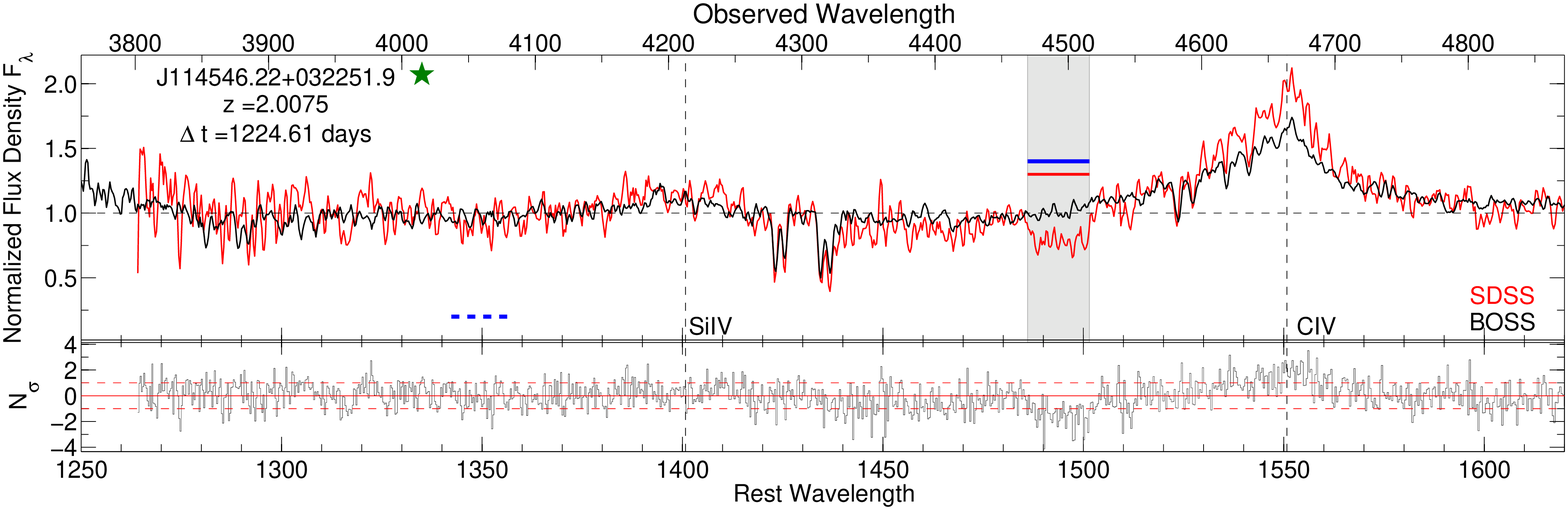}
\plotone{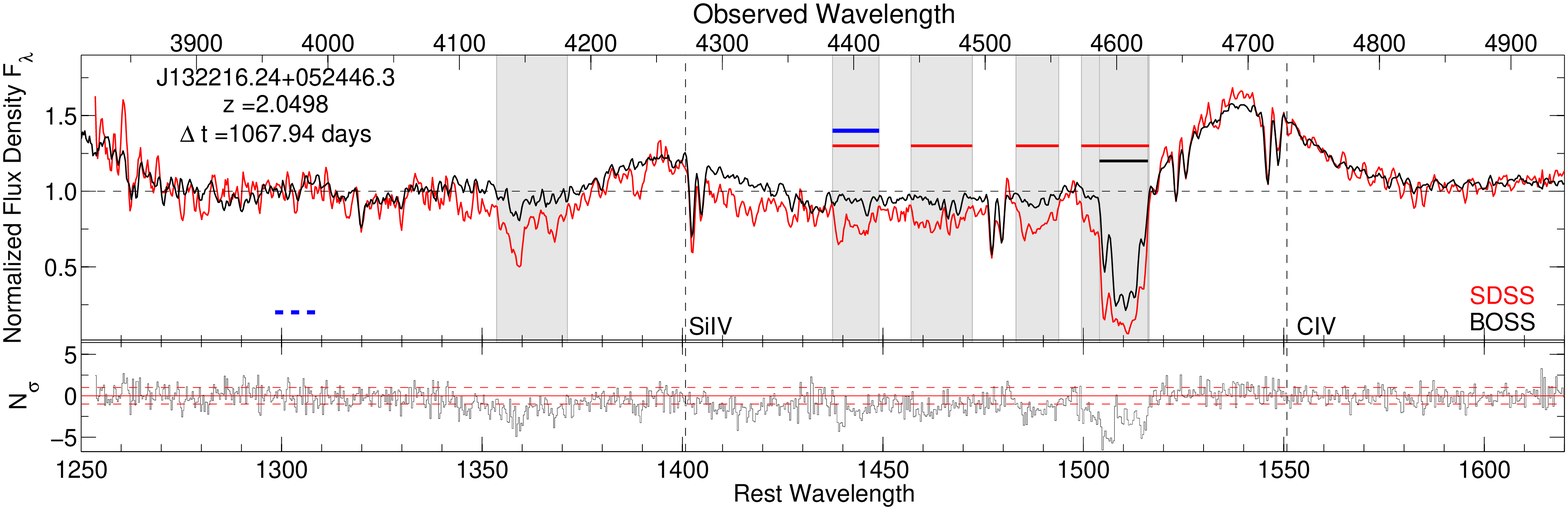}
\plotone{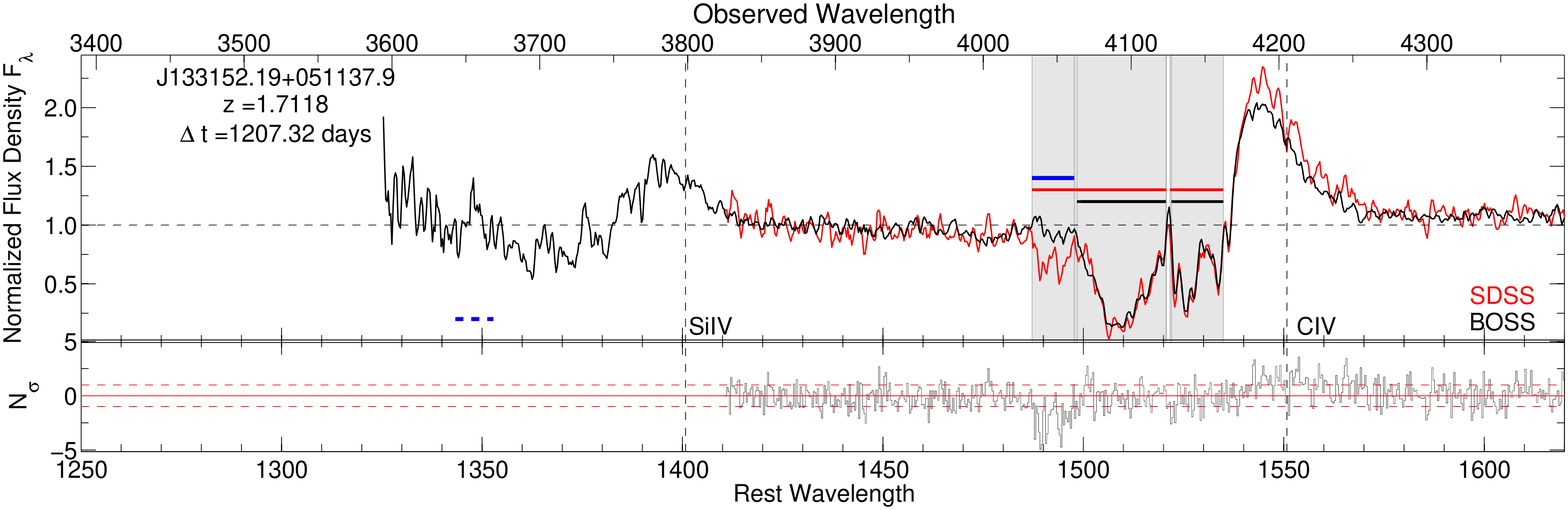}
\plotone{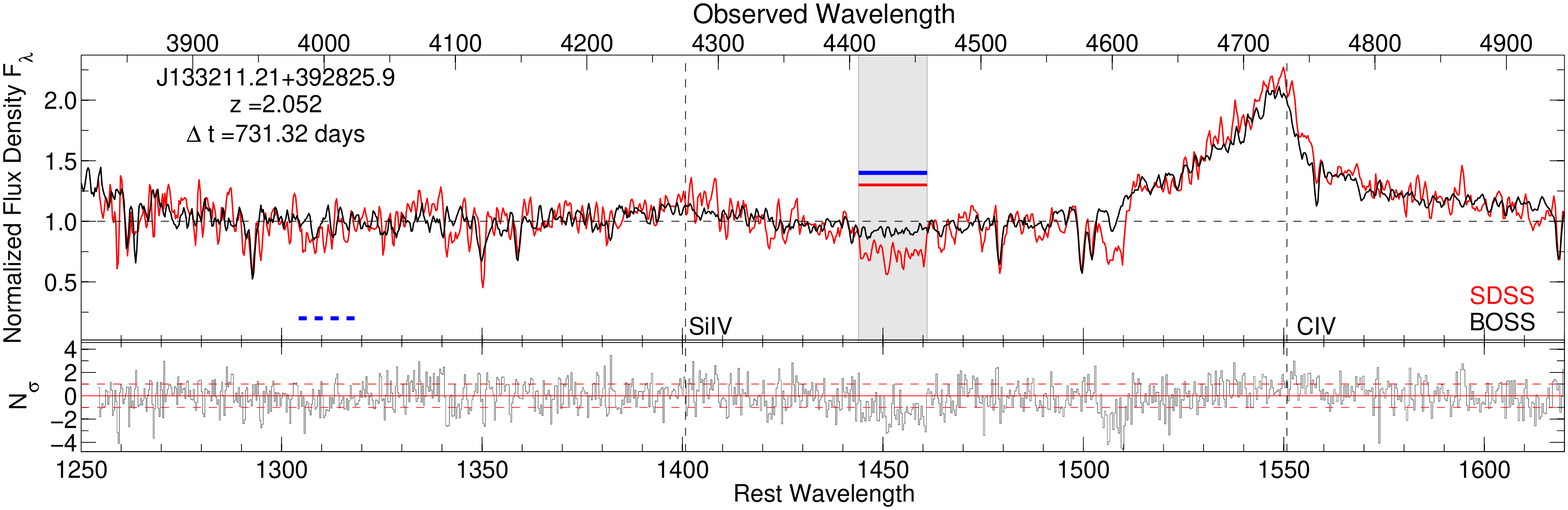}
\plotone{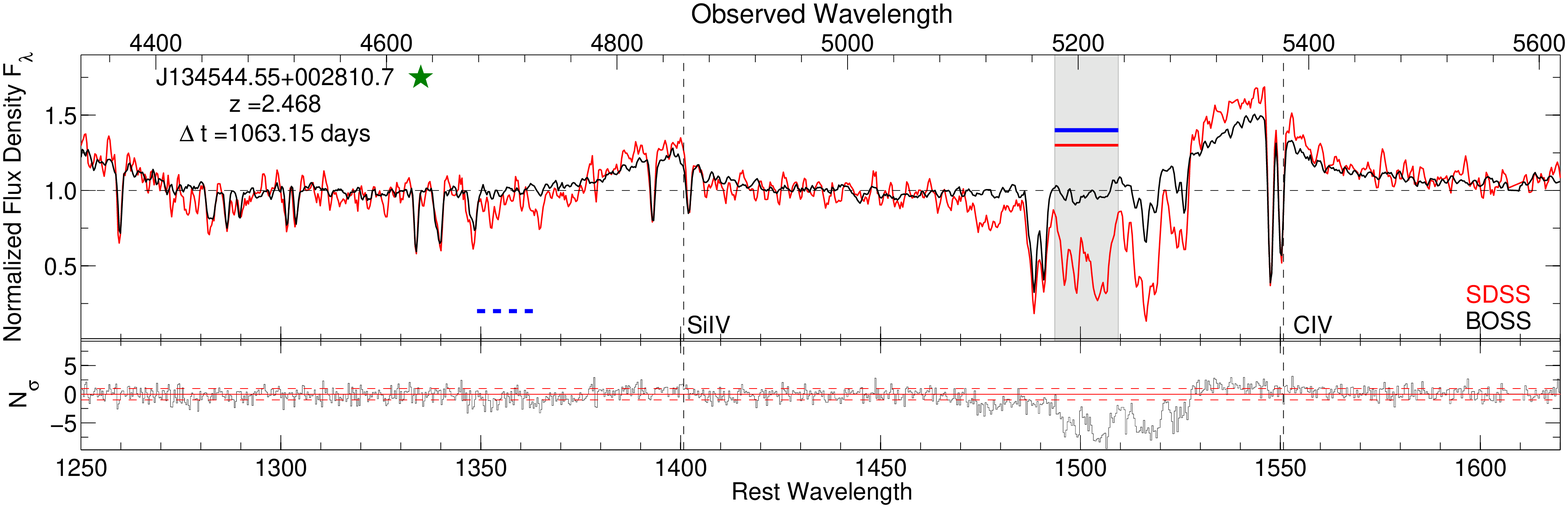}
\plotone{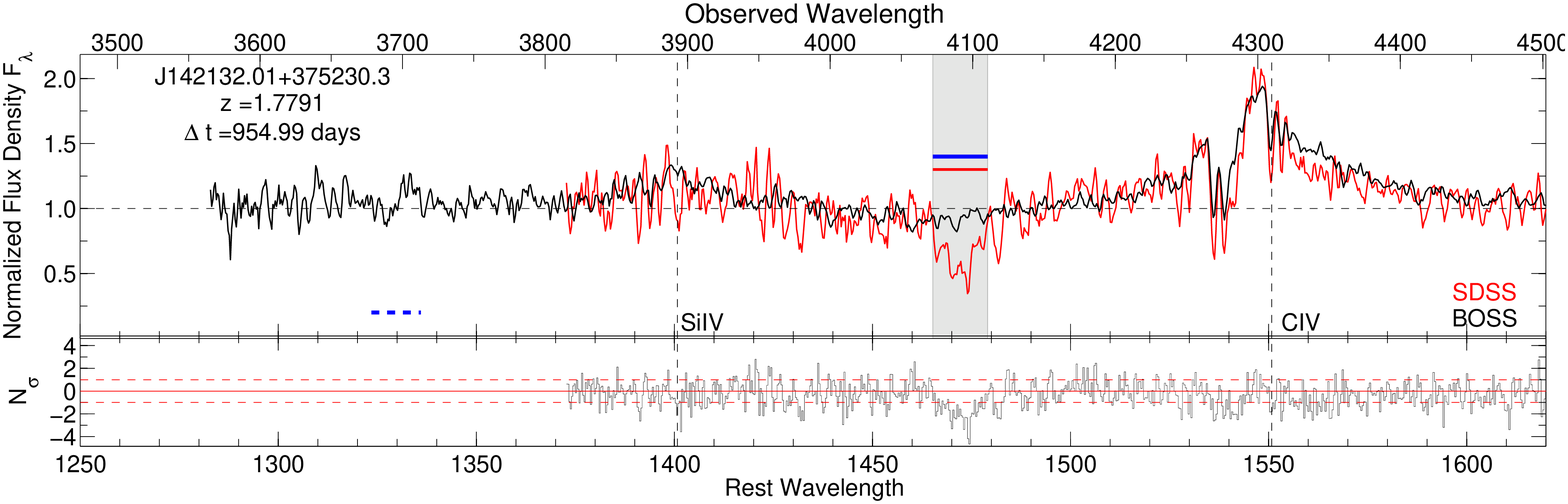}
\plotone{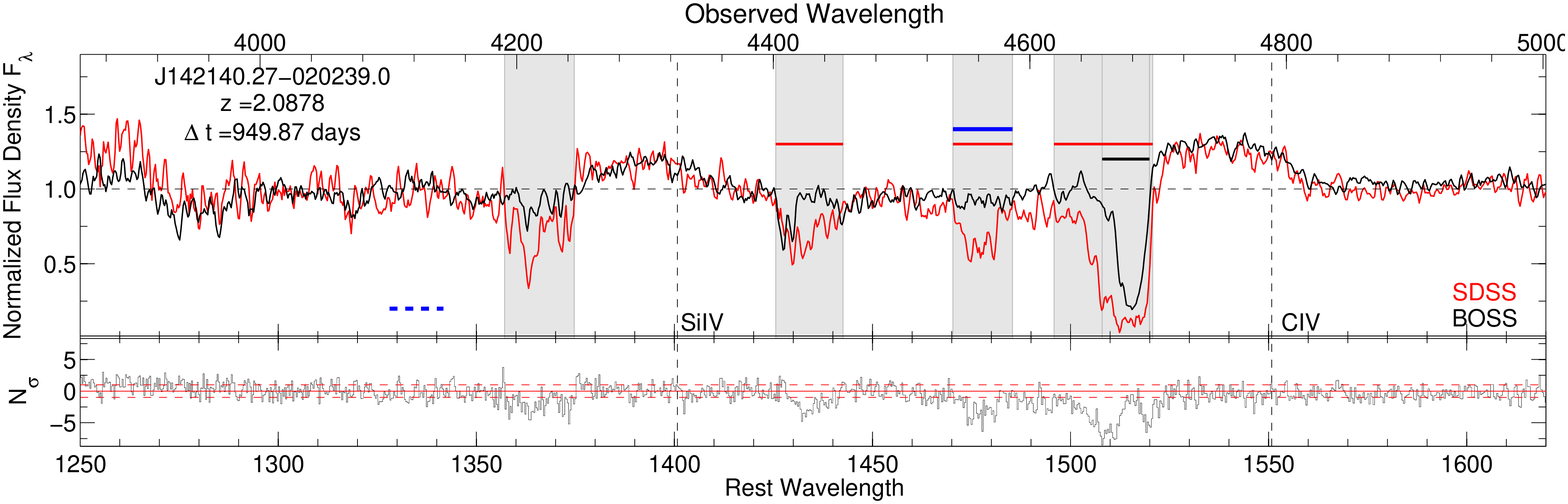}
\end{figure}

\begin{figure}
\epsscale{1.2}
\plotone{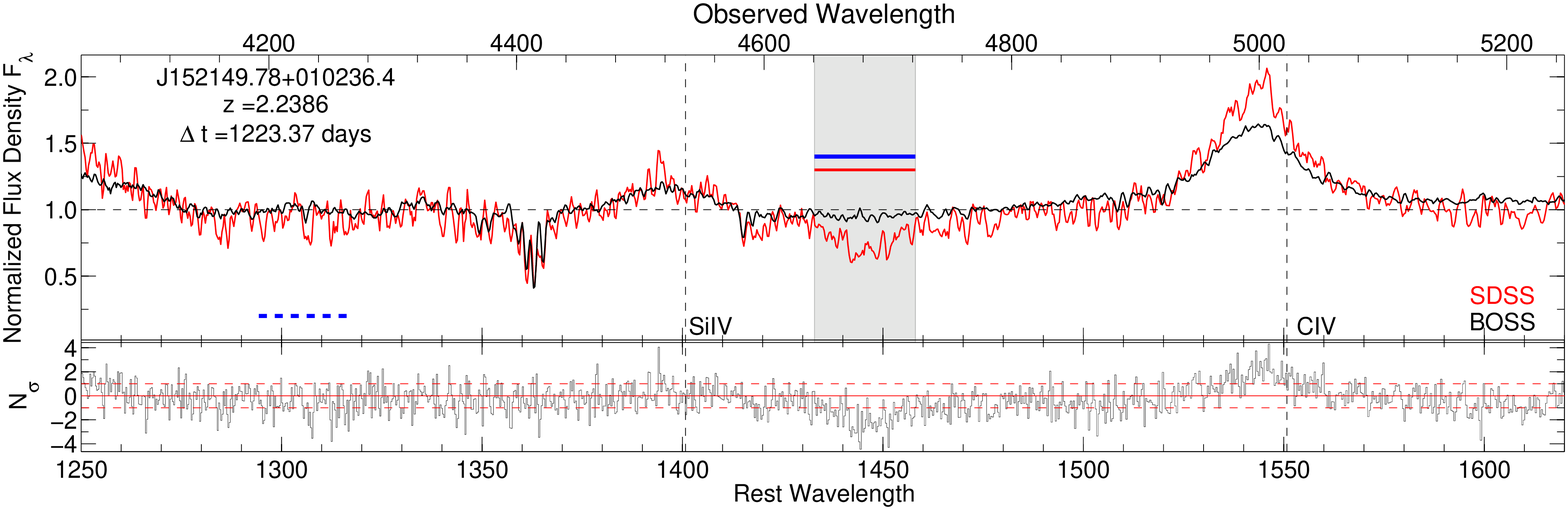}
\plotone{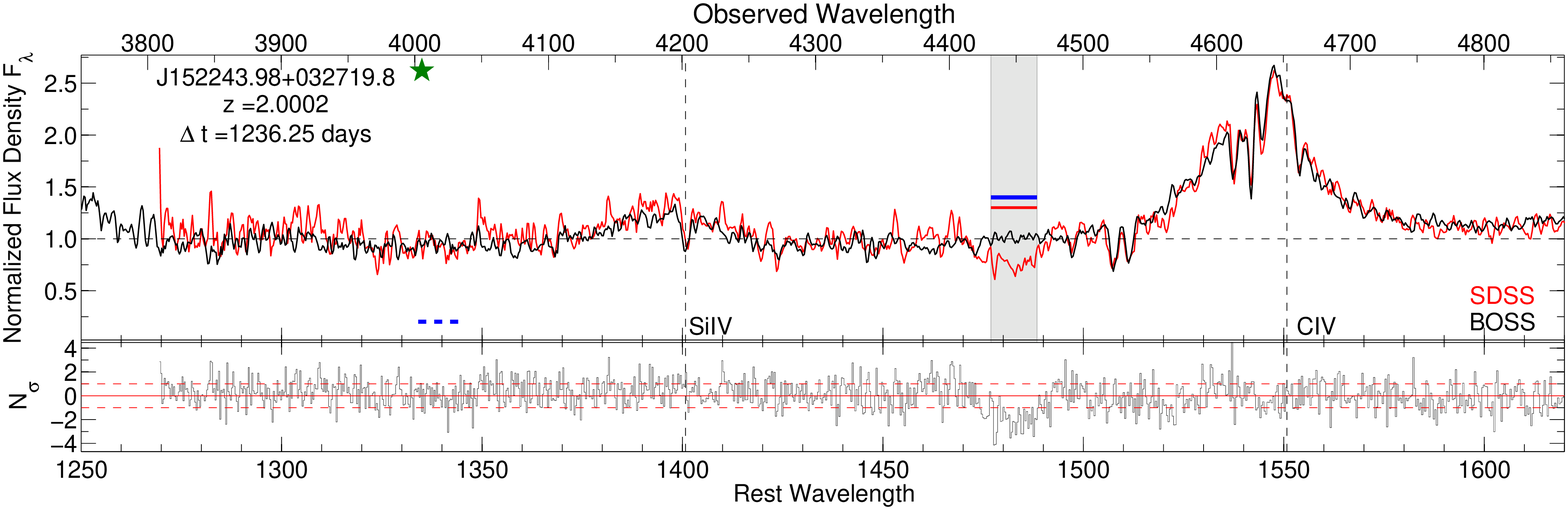}
\plotone{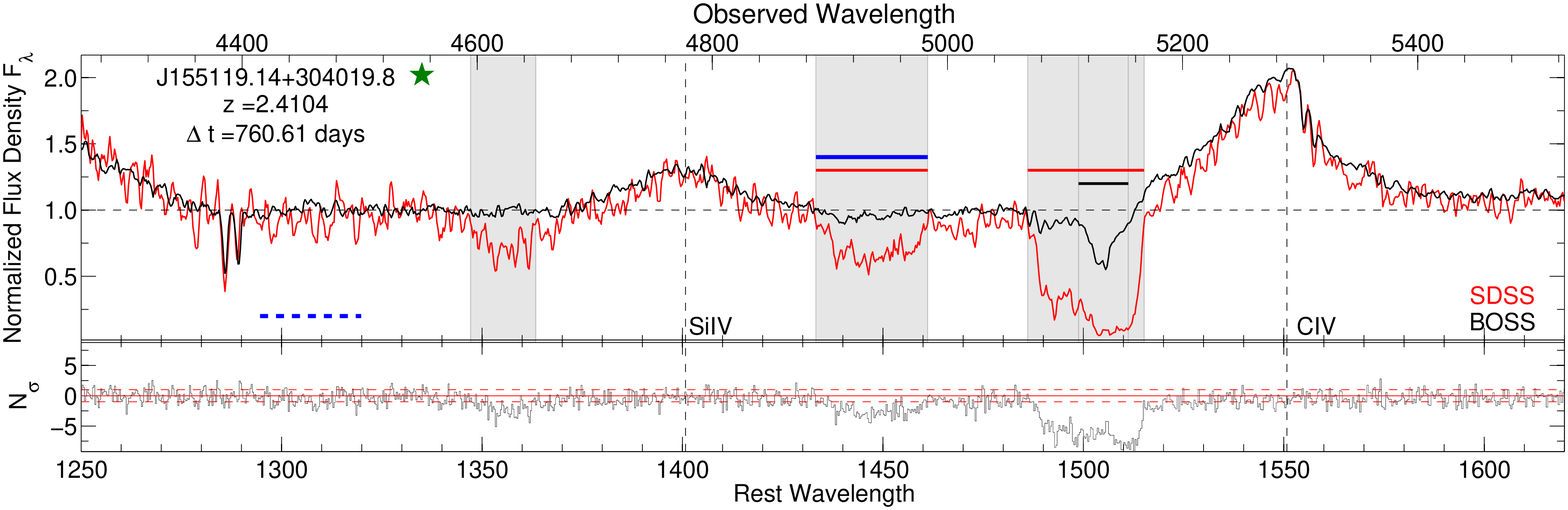}
\caption{Multi-epoch spectra of quasars with disappearing BAL troughs 
obtained from \hbox{SDSS-I/II} (red) and BOSS (black). The $x$-axes 
show observed (top) and rest-frame (bottom) wavelengths in 
$\mathrm{\AA}$. The dashed vertical lines show the emission-line rest 
wavelengths of  Si\,{\sc iv} $\lambda\lambda$\,1393,\,1402 $\mathrm{\AA}$ 
and C\,{\sc iv} $\lambda\lambda$\,1548,\,1550 $\mathrm{\AA}$. BALs are 
shown as shaded areas, and the red and black horizontal lines indicate 
BALs in the same color as the corresponding spectra. Solid blue bars show 
disappearing BAL troughs, and dashed blue bars show wavelengths of 
corresponding Si\,{\sc iv} velocities. Quasars with a green star next to their 
names are the ones that are identified as ``pristine'' examples of BAL 
disappearances (see \S\ref{selection}). 
The lower section of each panel shows the $N_{\sigma}$ values for SDSS vs. 
BOSS observations (black). The dashed horizontal lines show the 
\hbox{$\pm$ 1$\sigma$} level. }
\end{figure}

We have defined BAL disappearance following the formal definitions of 
BALs and mini-BALs given in \S\ref{measur}. However, in the BOSS spectra 
of some objects there is weak residual absorption that fails our formal  definition 
(see \S \ref{measur} and especially Equation~1) for being a BAL or mini-BAL. 
We define 11 of the 21 examples of disappearing C\,{\sc iv} BAL troughs as a 
``pristine'' sample that has no significant remaining absorption (NAL regions excluded).
These pristine sample objects are noted with stars in Figure~3.
Table~1 lists all observation epochs for the quasars with  
disappearing BAL troughs in addition to redshifts from \citet{hw10}, 
SDSS $i$-band magnitudes from \citet{schneider07}, and absolute 
$i$-band magnitudes from \citet{shen11}. The listed Plate-MJD-Fiber 
numbers are unique for each spectrum, and MJD $\geq$ 55176 indicates 
BOSS spectra. 
Table~2 lists the parameters of disappearing BAL troughs measured
in $S_1$ including observation epoch MJD, EW, $v_{\rm max}$, $v_{\rm min}$,
$f_{\rm deep}^{25}$, the rest-frame time interval between $S_1$ and $S_2$, 
and $\log(P_{\chi^2})$. 

To assess possible residual absorption remaining in the trough region of $S_2$, 
we set upper limits on EW using a $\chi^2$ fitting approach. We assume that the 
absorption remaining in $S_2$  has the same profile as that found in $S_1$, scaled 
by a constant  multiplicative factor, $\epsilon$, lying between 0 and 1. We first find 
the value of $\epsilon$ providing the best fit to the trough region in $S_2$ 
(spectral regions containing NALs as described in \S \ref{specific}  are excised in 
this analysis). We then set a 90\% confidence upper limit on $\epsilon$ by 
incrementing it until $\chi^2$ increases by 2.71 from the best-fit value 
\citep[see, e.g., \S15.6.5 of][]{press}, and we use the scaled profile with this upper 
limit on $\epsilon$ to compute an upper limit on EW. Our EW limits are typically 
$<1.3$~\AA\ with a median upper limit of 0.8~\AA. We note that our upper limits are 
conservative in that we do not allow for any narrowing of the absorption profile as it 
becomes weaker; analysis of our full BAL sample shows that such narrowing is typical.

We found that there are a few additional arguable cases of BAL disappearance that 
do not satisfy our formal definition for disappearing BAL troughs. Examples of such 
cases include J083817.00+295526.5, J092444.66$-$000924.1,  
J115244.20+030624.4, and  J131010.07+052558.8. 
The first three of these arguable cases arise because of difficulties 
in deciding if a BAL trough should be split into two distinct components 
(one of which disappears) or not; in such cases we follow the formal 
BAL definition criteria associated with Equation~1. The fourth case 
arises from a C\,{\sc iv} BAL lying close to the Si\,{\sc iv} emission line; 
in this case separation of continuum vs. line emission is difficult. 

We have investigated the Si\,{\sc iv} absorption at velocities corresponding 
to disappearing C\,{\sc iv} troughs. A total of 12 of 19 quasars have spectral 
coverage at the corresponding velocities in both the SDSS and the BOSS 
spectra. Two quasars (J004022.40+005939.6 and J081102.91+500724.2) show 
Si\,{\sc iv} absorption at the corresponding velocity, and this absorption 
disappeared in the BOSS spectra for both of these quasars (see Figure 3). 

We have investigated the Mg\,{\sc ii} and Al\,{\sc iii} regions of the 19 quasars with 
disappearing BAL troughs. All BOSS spectra have  full coverage of the Mg\,{\sc ii} 
region, as do SDSS spectra in all but  two cases (J004022.40+005939.6  and 
J134544.55+002810.7). None of the quasars with disappearing BAL troughs shows 
the presence of BALs in the Mg\,{\sc ii} or Al\,{\sc iii} regions in  their SDSS or BOSS 
spectra. Thus, certainly 17 and likely all 19 of these quasars are high-ionization BAL 
quasars. 

\begin{figure}[ht!]
\epsscale{0.57}
\plotone{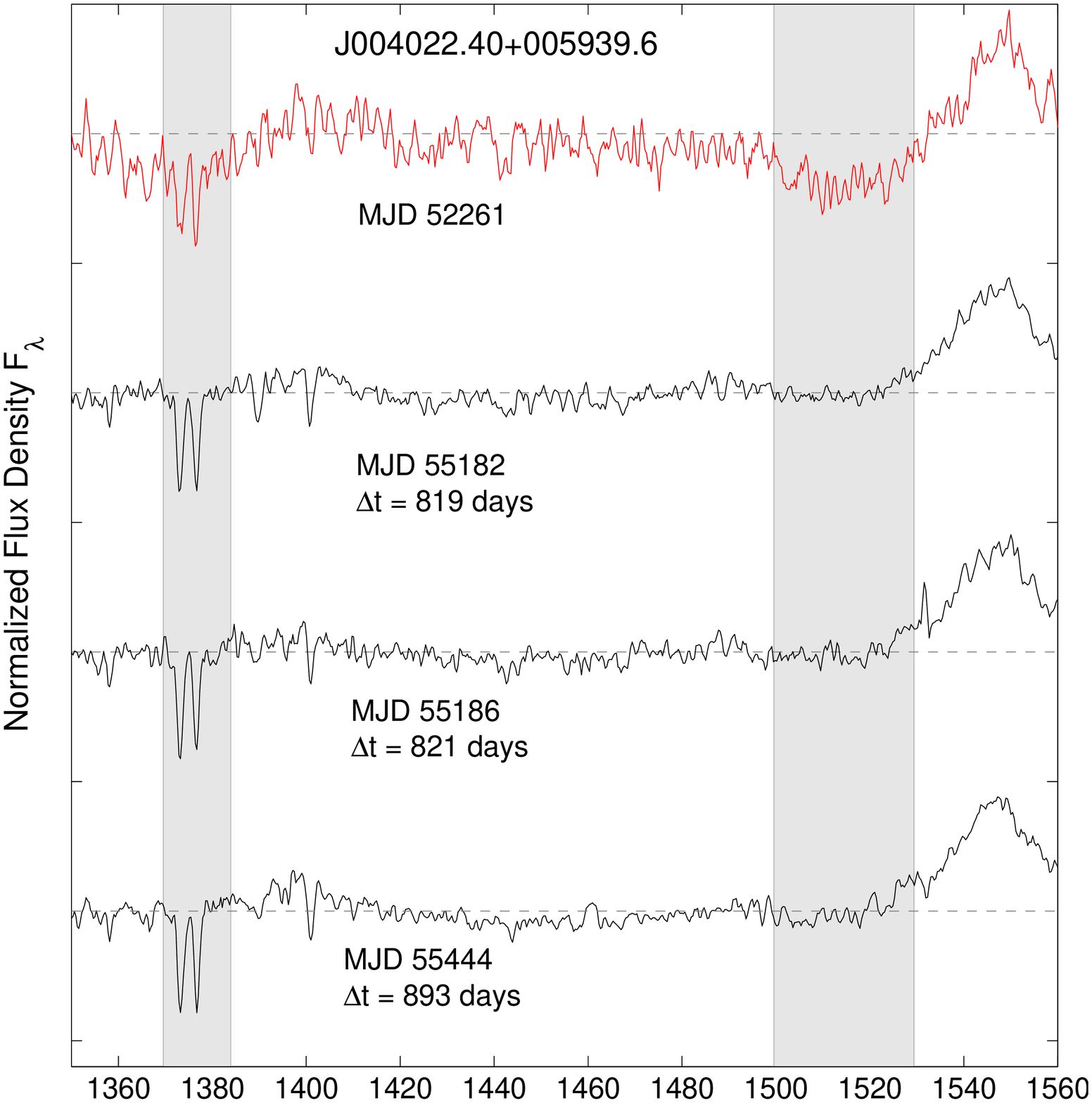}
\plotone{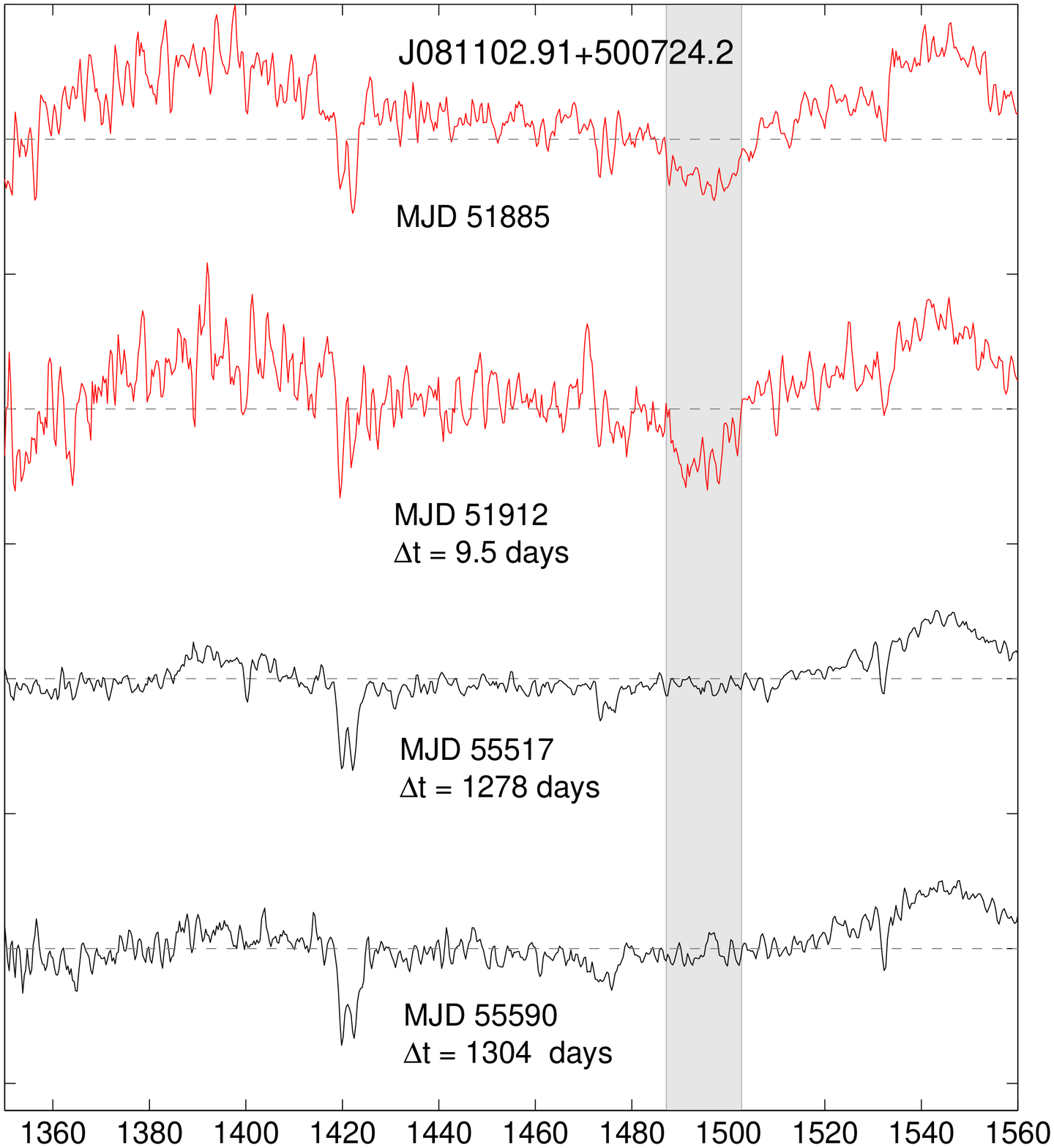}
\plotone{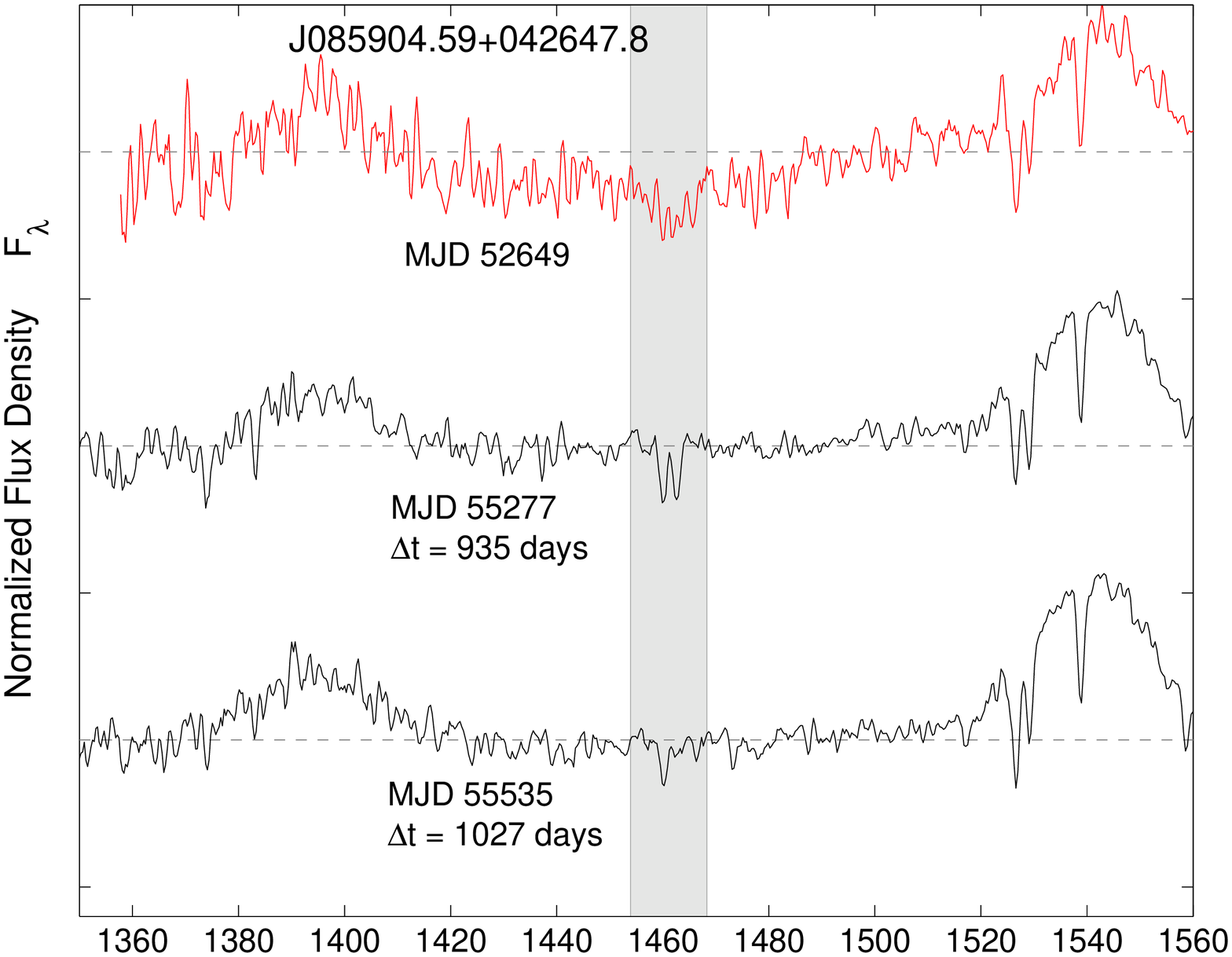}
\plotone{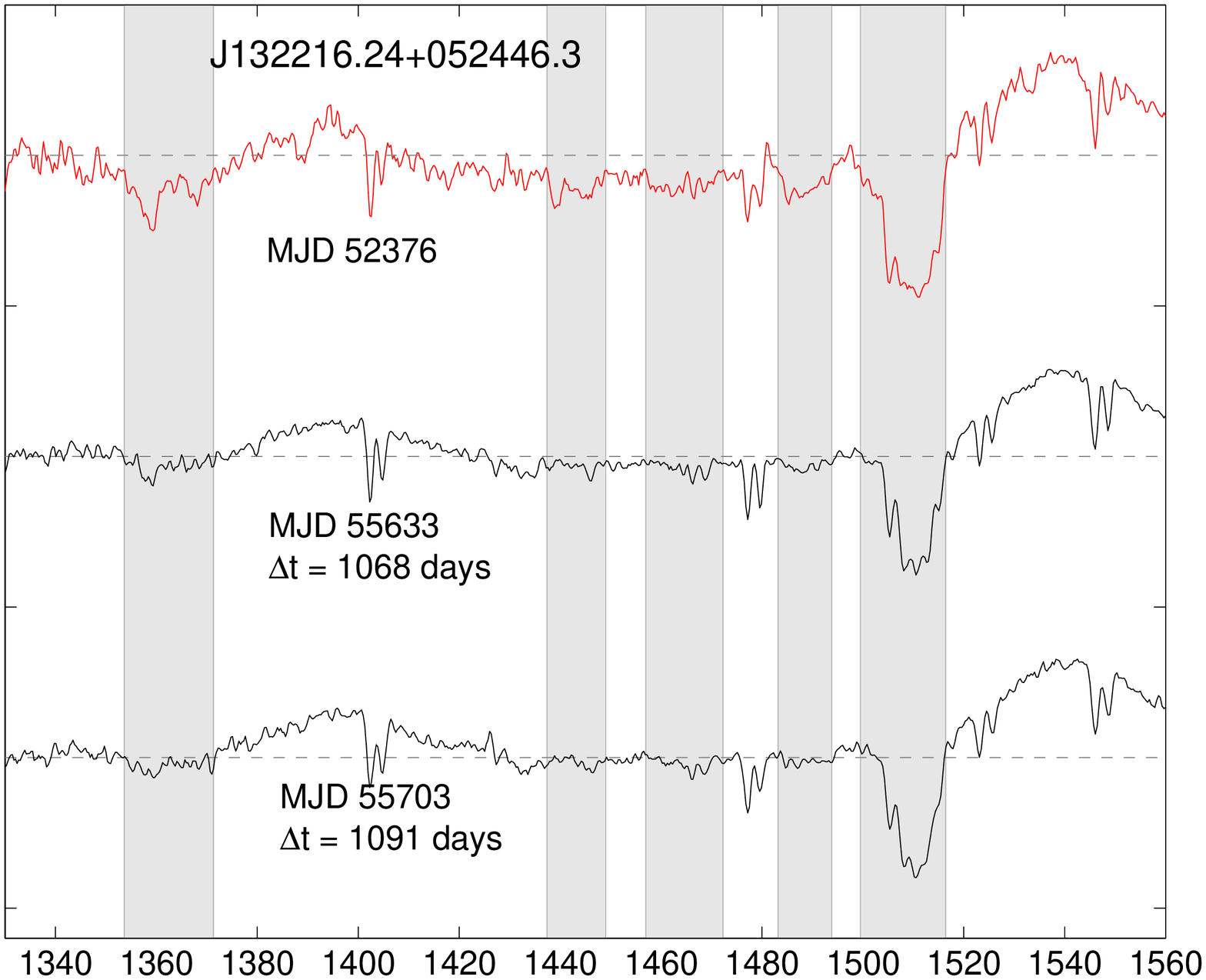}
\plotone{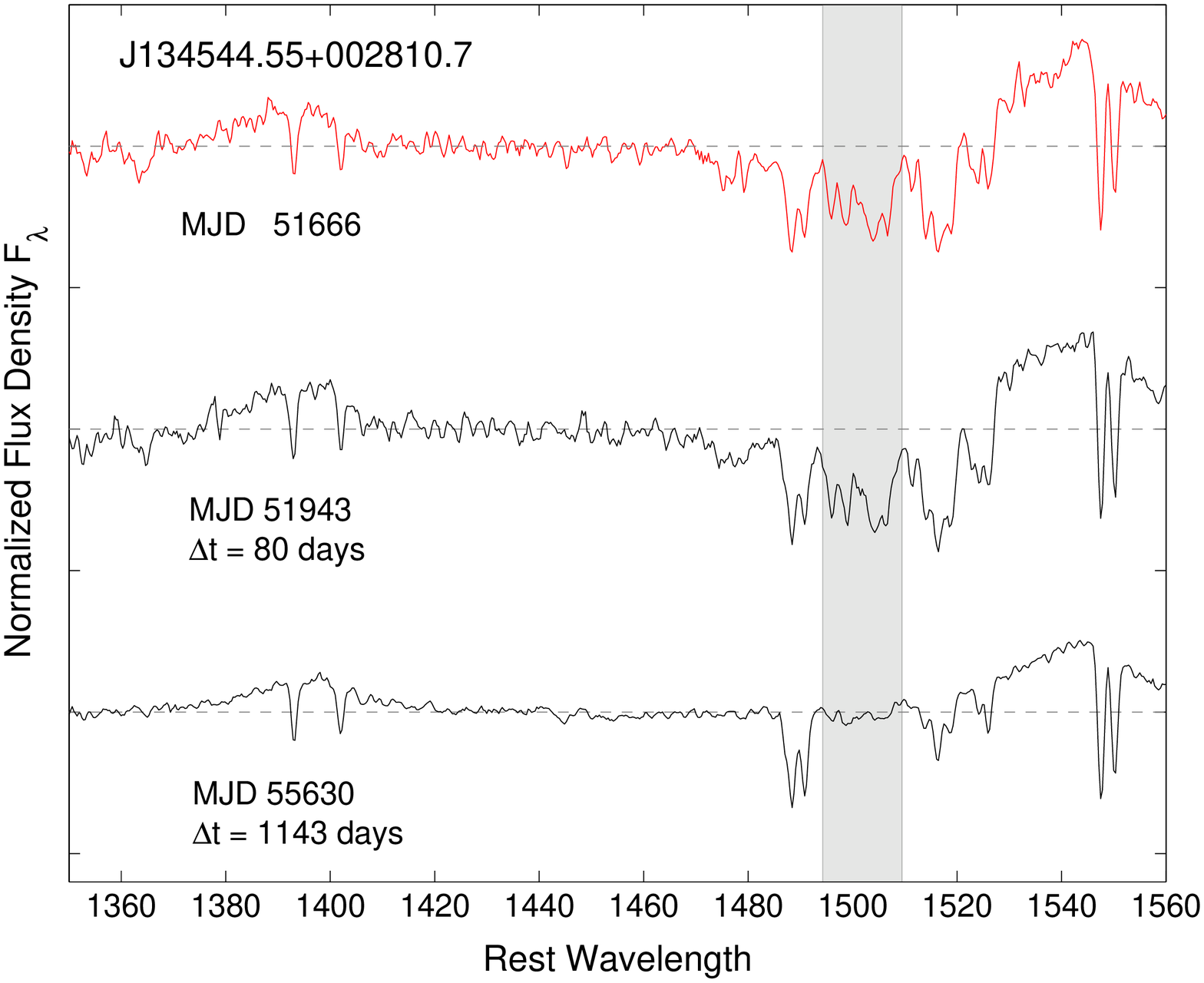}
\plotone{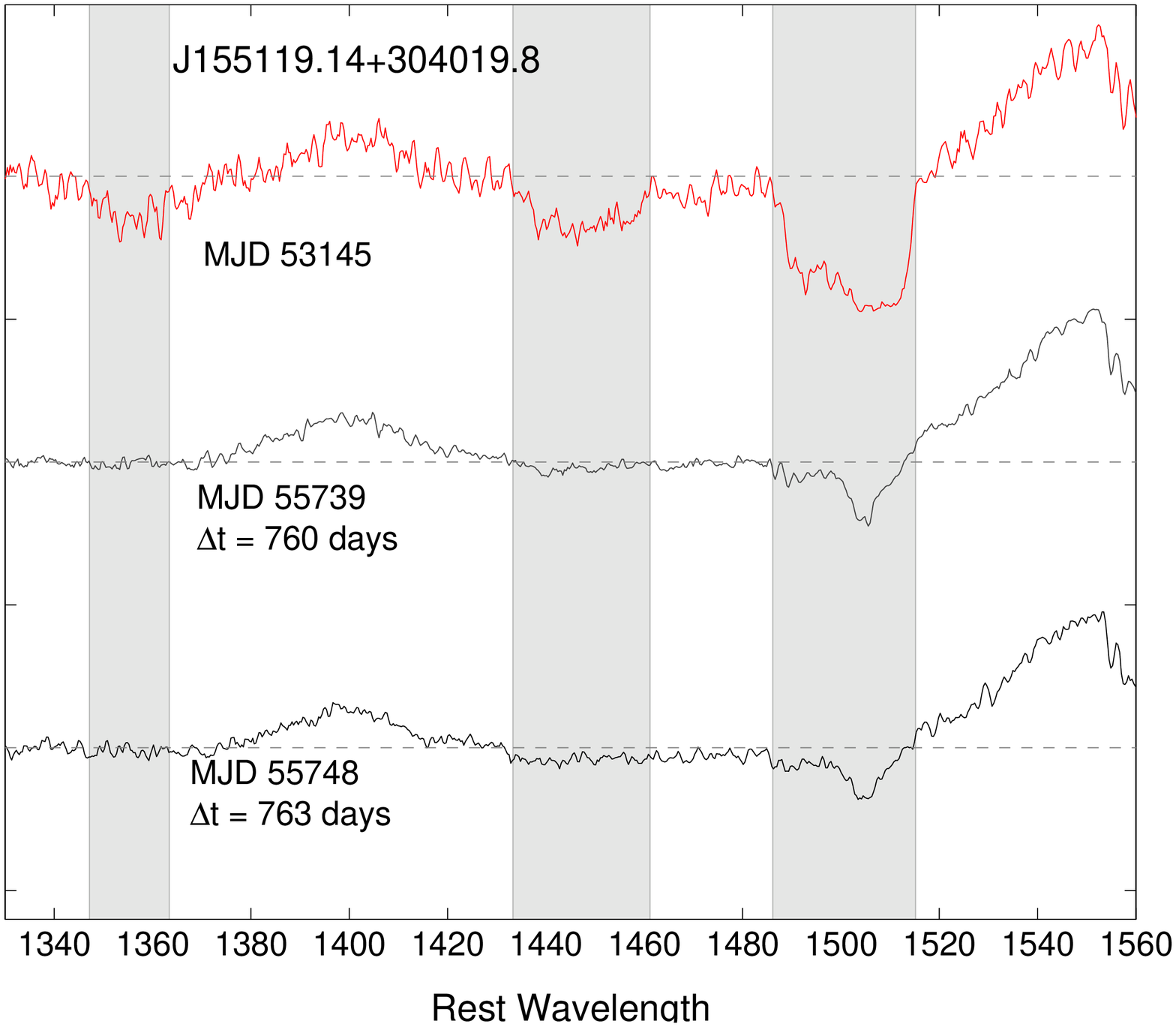}
\caption{Multi-epoch observations of the six quasars with disappearing 
BAL troughs that have more than one SDSS or BOSS spectrum. 
Continuum-normalized  SDSS (red) and BOSS (black) spectra are shown 
ordered by observation  date.  MJD values  for each observation  and the 
rest-frame time interval since the first observation are  indicated  under 
each spectrum. Horizontal dashed lines show continuum levels 
for  each spectrum, the thick marks on the $y$-axis show the zero level
for each spectrum, and shaded areas indicate BAL troughs. }
\label{fig4}
\end{figure}

Figure~\ref{fig4} shows the spectra of the six quasars with more than  two epochs of observation: 
J004022.40+005939.6,  J081102.91+500724.2,  J085904.59+042647.8, 
J132216.24+052446.3,   J134544.55+002810.7,  and J155119.14+304019.8. Note the 
consistency of the BOSS spectra across the multiple epochs, which confirms the BAL 
disappearance and the general robustness of our analyses.

\subsection{Notes on Specific Objects} \label{specific}

Here we describe the objects with remarkable properties or complex absorption 
spectra. Our descriptions supplement the objective approach adopted in 
\S \ref{selection} given the previously noted challenges sometimes associated with 
quantifying BAL troughs. 

\textit{J074650.59+182028.7:}
Both of the NAL doublets that are blended with the disappearing C\,{\sc iv}  BAL 
trough are identified as C\,{\sc iv} absorption in the BOSS NAL database \citep{lun}.

\textit{J085904.59+042647.8:}
The NAL doublet that is blended with the disappearing C\,{\sc iv} BAL trough is 
identified as C\,{\sc iv} absorption \citep{lun}. Figure~4 shows the three available 
epochs of observation for this quasar where the NAL appears to vary between two 
BOSS spectra (corresponding to a significance level of 95\%) separated  by 92~days 
in the rest frame. This apparent variability, combined with the location of this NAL 
within a BAL, suggest that the NAL is not an intervening absorption system at a lower 
redshift than the quasar but arises in gas outflowing from the quasar.

\textit{J093418.28+355508.3:}
The SDSS spectrum of this quasar shows an adjacent absorption feature  blueward 
(between rest-frame \hbox{1416--1421~\AA}) of the disappearing C\,{\sc iv} BAL trough. 
This feature could be a distinct mini-BAL trough or the continuation of a BAL trough 
that was truncated  in our formal approach owing to statistical fluctuations. Since this 
feature also disappears in the BOSS spectrum, its nature does not affect the interpretation 
of a BAL disappearance.

In the BOSS spectrum there is weak residual absorption that fails our formal 
definition (see \S \ref{measur} and especially Equation~1) for being a BAL or 
mini-BAL. 

\begin{figure}[h!]
\epsscale{1.15}
\plotone{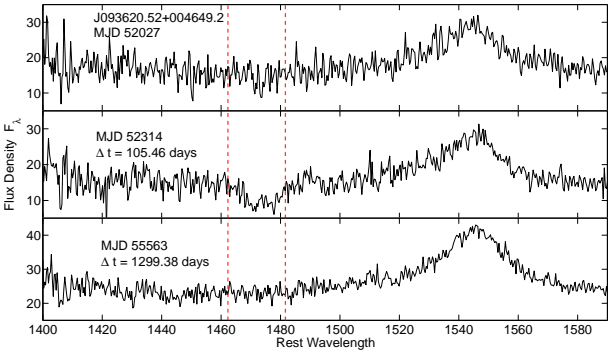}
\caption{Three-epoch observations of J093620.52+004649.2 that show the 
only reported example of a combined emergence and disappearance event 
for a BAL trough. The BAL trough of this object emerged between two SDSS 
observations \citep[top and middle panels;][]{lundgren07} and disappeared 
in the BOSS spectrum (bottom panel). The dashed red lines show the BAL 
trough region. } 
\label{fig5}
\end{figure}

\textit{J093620.52+004649.2:}
According to \citet{lundgren07}, the C\,{\sc iv} BAL trough of this object
emerged over a period of \hbox{$\leq105.6$~days} in the rest frame
between two SDSS observations; the possible C\,{\sc iv} absorption present
in the first SDSS epoch does not qualify as a BAL. We now find that this
same trough has disappeared over a period of $\leq 3.3$~yr, and no
remaining C\,{\sc iv} absorption of any kind is apparent in the BOSS
spectrum. Figure~\ref{fig5} shows the C\,{\sc iv} BAL trough region in all three
epochs. To our knowledge, this is the only reported example of a combined
emergence and disappearance event for a BAL trough.

\textit{J094806.58+045811.7:}
The disappearing BAL trough in this case is clearly distinct from the 
non-disappearing trough at higher velocity. The two troughs are separated 
by 22 pixels ($\approx 1500~\mathrm{km\,s^{-1}}$), 12 of which lie above 
the 90\% continuum threshold (see \S \ref{measur}). In fact, 6 pixels even 
lie above the continuum level.   

\textit{J104841.02+000042.8:}
The NAL doublet that is blended with the weak disappearing C\,{\sc iv}
BAL trough is identified as C\,{\sc iv} absorption \citep{lun}.

\textit{J112602.81+003418.2:}
In the BOSS spectrum there is weak residual absorption that fails our formal 
definition for being a BAL or mini-BAL. 

\textit{J132216.24+052446.3:}
In the BOSS spectrum there is weak residual absorption that fails our formal 
definition for being a BAL or mini-BAL. 

\textit{J133152.19+051137.9:}
The disappearing C\,{\sc iv} BAL trough is formally separated from
the adjacent BAL trough at lower outflow velocities according to our BAL 
identification criteria. However, it is possible that there is some degree of
physical connection between these two absorption systems. If they
are indeed significantly connected, this would undermine the evidence
for BAL disappearance in this quasar. It is notable, as apparent from
the $N_{\sigma}$ residuals in Figure 3, that the the lower velocity BAL trough
does not vary significantly as the higher velocity BAL trough disappears; this
is suggestive of a lack of physical connection between these two
systems. Further observations of this quasar are required for
clarification.

\textit{J133211.21+392825.9:}
In the BOSS spectrum there is weak residual absorption that fails our formal 
definition for being a BAL or mini-BAL.

\textit{J134544.55+002810.7:}
This is the most complex, as well as the strongest, absorption system
in our sample of disappearing BALs. In the SDSS spectrum, the disappearing
BAL trough appears to be attached to a strong higher velocity C\,{\sc iv} NAL
doublet \citep[see][]{lun} as well as a lower velocity mini-BAL. After the
disappearance of the BAL trough, the NAL remains while the mini-BAL
transforms to a NAL. Even if we consider the NAL, BAL and mini-BAL
together as a connected BAL complex, the complex would still be listed
as a disappearing BAL (since the only remaining features in the BOSS
spectrum are NALs; see \S3.3). We also note that there is another
potentially connected mini-BAL at \hbox{1470--1485~\AA} in the
SDSS spectrum; this mini-BAL disappears in the BOSS spectrum.

\textit{J142132.01+375230.3:}
In the BOSS spectrum there is weak residual absorption that fails our formal 
definition for being a BAL or mini-BAL.

\textit{J142140.27$-$020239.0:}
In the BOSS spectrum there is weak residual absorption that fails our formal 
definition for being a BAL or mini-BAL.

\textit{J152149.78+010236.4:}
In the BOSS spectrum there is weak residual absorption that fails our formal 
definition for being a BAL or mini-BAL.

\section{Statistical Properties of Disappearing BALs}

\subsection{How Common is BAL Disappearance?} \label{common}

The disappearance of BAL troughs is expected to be a rare event 
\citep[e.g.,][]{gibson08,hall11}. We have investigated 925 distinct BAL troughs in 
582 quasars observed over \hbox{1.1--3.9~yr} in the rest frame. From this sample, 
we have identified $N_{\rm dt} = 21$ disappearing BAL troughs in $N_{\rm qdt} = 19$ 
distinct quasars. Thus, on the observed timescales, $f_{\rm disapp} = 2.3^{+0.6}_{-0.5}$\% 
(i.e., 21/925) of BAL troughs disappear, and $f_{\rm quasar} = 3.3^{+0.9}_{-0.7}$\% 
(i.e., 19/582) of BAL quasars show a disappearing trough \citep[quoted error bars are 
at $1\sigma$ confidence following][]{gehrel}. 
If we consider only the pristine-sample objects defined in \S \ref{selection}, the observed 
fraction of disappearing C\,{\sc iv} BAL troughs is $f_{\rm disapp} = 1.2^{+0.6}_{-0.5}$\% 
(i.e., 11/925) and the fraction of BAL quasars showing a disappearing trough is 
$f_{\rm quasar} = 1.9^{+0.8}_{-0.6}$\% (i.e., 11/582).
Table~3 lists the stated percentages above and the other quantities introduced below
both for the standard sample and the pristine sample.  

One potential implication of the observed frequency of disappearing C\,{\sc iv} BAL 
troughs is a relatively short average trough rest-frame lifetime
\hbox{$\bar{t}_{\rm trough}\approx\langle\Delta t_{\rm max}\rangle/f_{\rm disapp}
= 109_{-22}^{+31} {\rm~yr} $}, where $\langle\Delta t_{\rm max}\rangle$ is the 
average of the maximum time between observation epochs in a sample of BAL 
quasars (2.5~yr in this case) and $f_{\rm disapp}$ is the fraction of troughs that 
disappear over that time (also see Table~3). Here we state the average trough 
lifetime and average maximum observation time with different notations indicating 
that $\bar{t}_{\rm trough}$ is inferred, whereas $\langle\Delta t_{\rm max}\rangle$ 
is measured.
Note that by lifetime we mean the time over which a trough is seen along our line 
of sight and not necessarily the lifetime of the gas clouds responsible for the absorption. 
The BAL phenomenon can last longer than $\bar{t}_{\rm trough}$ if troughs come and 
go along our line of sight within BAL quasars. If many troughs 
have extremely long lifetimes then the above is only a lower limit on the true
$\bar{t}_{\rm trough}$.  In that case, the fraction $f_{\rm short}$ of relatively 
short-lived troughs that dominate the parent population of disappearing troughs 
must have lifetimes of only $f_{\rm short}\,\bar{t}_{\rm trough}$.  Thus, we conclude 
that a significant fraction of BAL troughs have average lifetimes of a century or less.

\begin{figure}[h!]
\epsscale{1.2}
\plotone{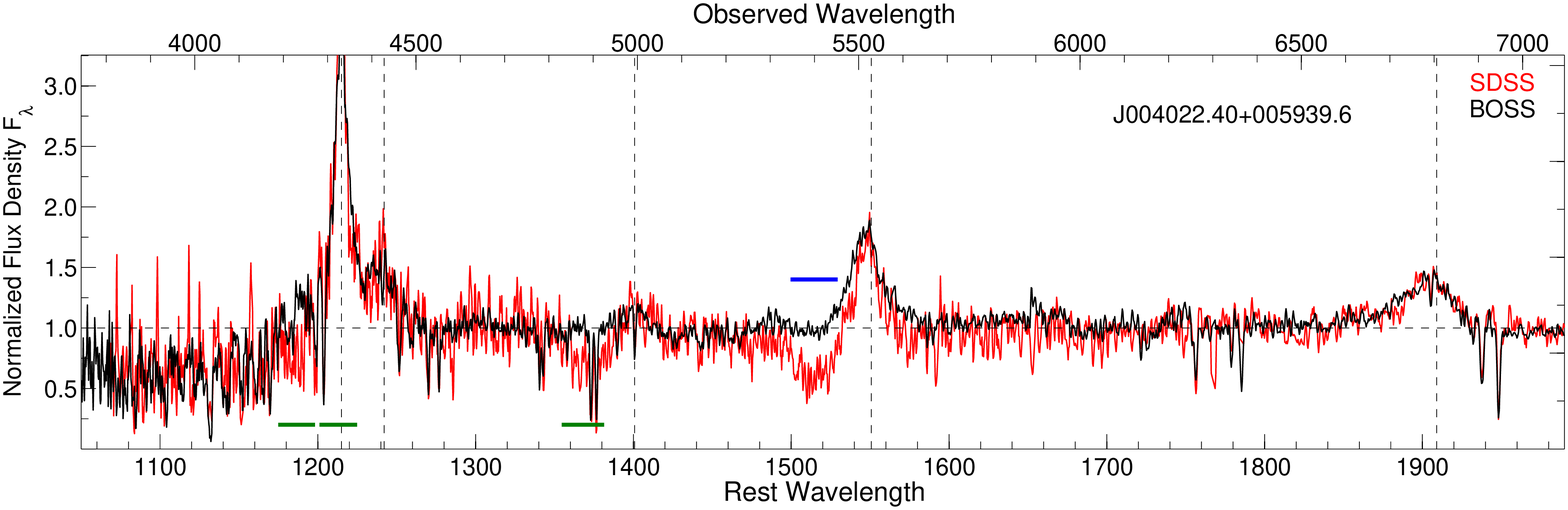}
\plotone{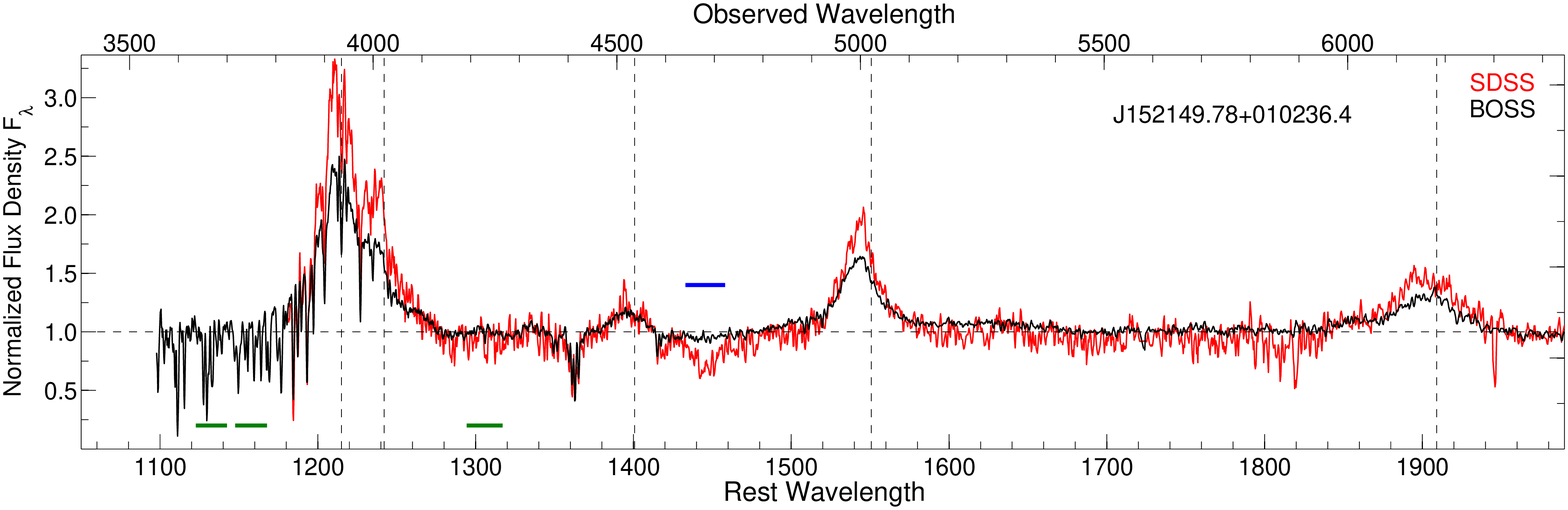}
\caption{BAL quasars J004022.40+005939.6, J133211.21+392825.9, and 
J152149.78+010236.4 that transformed to non-BAL quasars after the 
disappearance of the C\,{\sc iv} BAL trough (marked with a blue bar). Green 
bars under each spectrum indicate the velocities in the Ly$\alpha$, N\,{\sc v}, 
and Si\,{\sc iv} regions corresponding to the disappearing C\,{\sc iv} trough. 
Vertical dashed lines indicate the emission lines for Ly$\alpha$, N\,{\sc v}, 
Si\,{\sc iv}, C\,{\sc iv}, and  C\,{\sc iii}, respectively, from left to right.} 
\label{fig6}
\end{figure}

Notably, our sample includes $N_{\rm transform} = 10$ examples of objects that 
have apparently transformed from BAL to non-BAL quasars; these are denoted in 
Table~1. BOSS spectra of these objects demonstrate that their C\,{\sc iv} and 
Si\,{\sc iv} regions do not have any remaining BAL (or mini-BAL) troughs.\footnote{Two 
of these objects, J114546.22+032251.9 and J152149.78+010236.4, do have strong 
multiple NAL systems in their BOSS spectra that do not overlap in velocity with the 
disappearing C\,{\sc iv} BAL trough.}
Only two of these 10 objects (J004022.40+005939.6 and J152149.78+010236.4) 
have  BOSS spectral coverage of the corresponding Ly$\alpha$ and N\,{\sc v} BAL 
transitions as well, and visual inspection confirms that these spectra do not show 
any remaining BAL or mini-BAL troughs in Ly$\alpha$ or N\,{\sc v} 
(see Figure~\ref{fig6}).\footnote{The BOSS spectrum of J133211.21+392825.9 shows 
intervening absorption from a damped Ly$\alpha$ absorber that is not related to 
the BAL phenomenon. The intervening nature of this absorption is supported by 
its narrow width and symmetric shape, as well as by the fact that the maximum 
depth of the line is consistent with zero flux.}
As we are not aware of any BAL quasars that show convincing Ly$\alpha$ 
or N\,{\sc v} absorption without also showing C\,{\sc iv} or Si\,{\sc iv} absorption, 
we consider the lack of remaining absorption in the latter transitions alone to 
be credible evidence of transformation from a BAL to a  
non-BAL quasar.\footnote{One {\it possible\/} counterexample is the remarkable 
quasar PDS~456 which may show only broad Ly$\alpha$ absorption \citep{obrien05}. 
However,  the identification of the single broad absorption feature found in the 
spectrum of this quasar is not clear; e.g., it could be from highly 
blueshifted C\,{\sc iv}.} 
The one qualification here is that it is possible, in principle, that 
some of these objects could have transitioned to a very 
high-ionization BAL state like that seen in SBS~1542+541 
\citep{telfer98}; SBS~1542+541 shows strong O\,{\sc vi} absorption 
but only weak or no absorption in transitions like C\,{\sc iv},
Si\,{\sc iv}, and N\,{\sc v}. 
However, we consider this an unlikely possibility because
such very high-ionization BAL quasars appear to be rare
based on a visual search of the spectra of high-redshift SDSS DR7
catalog \citep{schneider10} quasars (by P.B.H.) and on their absence as contaminants
in searches for damped Ly$\alpha$ absorption systems
in the same catalog (J.X. Prochaska 2011, personal communication).
The fraction of BAL quasars transforming to non-BAL quasars is   
$f_{\rm transform} = 1.7^{+0.7}_{-0.5}$\% (i.e., 10/582) on \hbox{1.1--3.9~yr}  
timescales. These are the first reported examples of BAL to 
non-BAL quasar transformations; the four previously reported BAL
disappearance events described in \S\ref{intro} did not involve 
transformations from BAL to non-BAL quasar status (i.e., troughs 
from other ions or additional C\,{\sc iv} troughs remained in the 
spectra).

These BAL to non-BAL quasar transformations set a lower limit on the
lifetime of the ultraviolet BAL phenomenon along our line of sight of
\hbox{$\bar{t}_{\rm BAL}\simeq \langle\Delta t_{\rm max}\rangle/
f_{\rm transform} = 150_{-50}^{+60} {\rm~yr}$}.
However, defining the lifetime of the BAL phase along our our line of sight
is more complicated than defining a trough lifetime along our line of sight.
For example, if a BAL quasar has its only trough disappear but then has a 
different trough appear a year later, should this event be considered separate 
BAL phases or one phase with patchy absorption? Monitoring the strong X-ray 
absorption which is also characteristic of BAL quasars \citep[e.g.,][]{gal02,gal06} 
would improve our understanding of how connected UV absorption variability is 
to absorption variability at other wavelengths, and thus to variability between 
BAL and non-BAL phases along a given line of sight, all of which can help test 
BAL outflow models.

\subsection{Luminosities, Black-Hole Masses, Reddening, and 
Radio Properties of Quasars with Disappearing BAL Troughs} \label{qsos}

\begin{figure}[ht!]
\epsscale{1.2}
\plotone{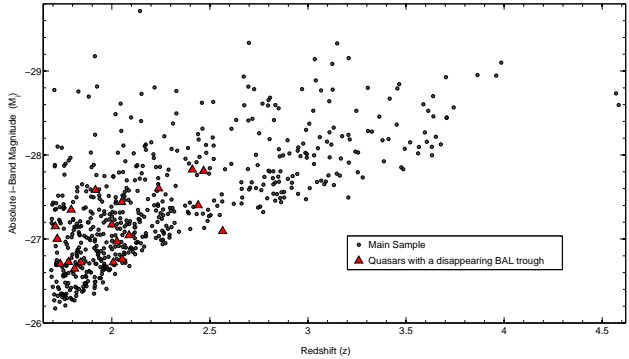}
\caption{Redshifts vs. absolute $i$-band magnitudes of main-sample 
quasars in this study (black solid circles) and quasars with disappearing 
C\,{\sc iv} BAL troughs (red triangles). Redshifts are from \citet{hw10},  
and $M_{i}$ values are from \citet{shen11}.} 
\label{fig7}
\end{figure}

In Figure~\ref{fig7}, we compare our main-sample quasars to quasars with 
disappearing troughs in  a plot of redshift versus absolute $i$-band 
magnitude ($M_i$).  We physically expect that, since the sampled  
rest-frame timescales are shortened toward higher redshifts, less 
BAL variability will be seen with increasing redshift. Therefore, we 
applied the two-sample Kolmogorov-Smirnov (K-S) test to compare 
$M_i$ distributions for quasars with disappearing BAL troughs and 
the main-sample quasars sampled on similar rest-frame timescales 
(i.e., $z \leq 2.6$). We find no evidence that BAL quasars with 
disappearing troughs have exceptional $M_i$ values compared to 
the main-sample quasars with similar redshifts (K-S probability of 
68\%).
In addition, we compared SMBH mass estimates from \citet{shen11} 
for quasars with disappearing BAL troughs and our main-sample 
quasars.  The quasars with disappearing C\,{\sc iv} BAL troughs do 
not show any remarkable  inconsistency from the main sample (K-S 
probability of 9\% for $z \leq 2.6$ quasars).  Thus, the disappearing 
trough phenomenon appears to be generally present throughout the 
BAL quasar population. 

We have also investigated if our quasars with disappearing C\,{\sc iv}
BAL troughs show any difference in intrinsic reddening  from the main 
sample. Following \citetalias{gibson09}, we have calculated a basic 
``reddening parameter'', defined as the ratio of the continuum flux densities at 
1400~\AA\ and 2500~\AA. We do not detect any difference in the 
distributions of this parameter for the quasars with disappearing 
C\,{\sc iv} BAL troughs and the main sample.

The FIRST survey has detected radio emission from two of the 
quasars with disappearing troughs. J004022.40+005939.6  is 
radio intermediate with \hbox{$R = 60.7$} and  J081102.91+500724.2 
is radio loud with \hbox{$R = 223.8$}. The quasars without  detected 
radio emission have \hbox{$R < 5$} and are thus radio quiet. The BAL
variability of radio-loud BAL quasars is not well understood and is just 
now being studied systematically \citep[e.g.,][]{miller12}. Our results for 
J004022.40+005939.6 and J081102.91+500724.2 demonstrate that 
BAL disappearance is a phenomenon of both radio-loud and 
radio-quiet quasars.

\subsection{EWs, Depths, Velocities, and  Widths of Disappearing BAL Troughs} 
\label{ews}

\begin{figure}[h!]
\epsscale{1.15}
\plotone{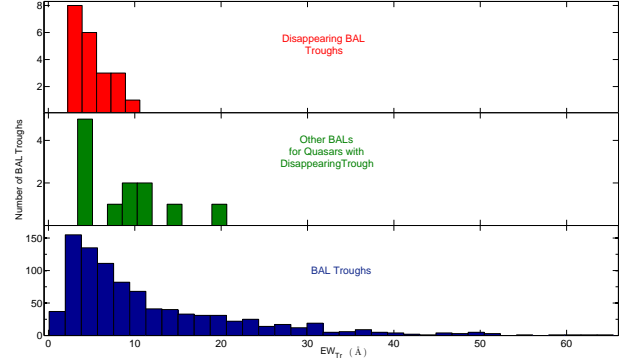}
\caption{EW distributions for disappearing  C\,{\sc iv} BAL troughs (upper 
panel, red), the other BAL troughs present in quasars that show one 
disappearing trough (middle panel, green), and all 925 distinct C\,{\sc iv} 
BAL troughs in the main sample (lower panel, blue).}
\label{fig8}
\end{figure}

We have investigated the characteristics of disappearing C\,{\sc iv} BAL 
troughs by comparing them with non-disappearing ones. 
Figure~\ref{fig8} shows a comparison of the EW distributions for disappearing 
BAL troughs, the other BAL troughs present in quasars that show one 
disappearing trough (i.e., those that do not disappear; see 
\S\ref{connection} for further discussion), and all 925 
distinct BAL troughs in the main sample. In order to be consistent, we 
measured the BAL-trough parameters from the observations obtained 
at epoch $t_1$ (i.e., the latest observation from the SDSS). 
The K-S test results show that there is only a 0.09\% chance of 
consistency between the EW distributions of disappearing BAL troughs 
and all main-sample BAL troughs. Based on this result and Figure~8, 
it appears that BAL disappearance tends to occur for weak or 
moderate-strength absorption troughs but not the strongest ones. In 
particular, no troughs with EW~$>12$~\AA\ disappeared. 
Furthermore, the K-S test results comparing the main-sample 
C\,{\sc iv} BAL troughs and the additional non-disappearing C\,{\sc iv} 
BAL troughs defined in \S \ref{connection} show no evidence for 
inconsistency (K-S probability of 56.9\%). 

\begin{figure}[h!]
\epsscale{1.15}
\plotone{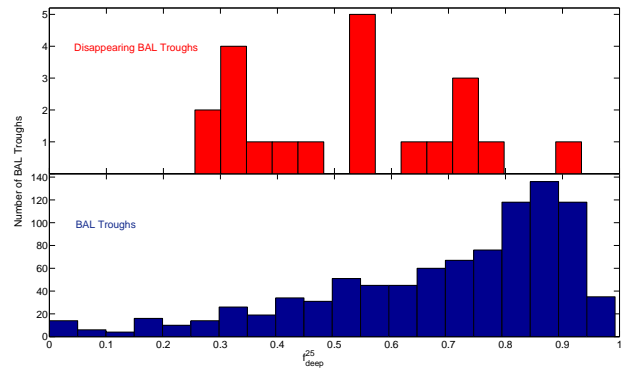}
\caption{Distribution of $f_{\rm deep}^{25}$ for disappearing C\,{\sc iv}
BAL troughs (upper panel) and for all 925 distinct C\,{\sc iv} BAL troughs in 
the main sample (lower panel).}
\label{fig9}
\end{figure}

Figure~\ref{fig9} shows the distributions of the BAL-trough depth parameter, 
$f_{\rm deep}^{25}$, for disappearing BAL troughs and for all 925 distinct 
BAL troughs in the main sample. The K-S test shows that these two 
distributions only have a 0.03\% chance of consistency. The BAL 
troughs that disappear  are shallower than BAL troughs in general, 
although some fairly deep BAL troughs do disappear. 

Figures~10a and 10b show the distributions of $v_{\rm max}$ and 
$v_{\rm min}$ (see \S \ref{measur}) for disappearing and all 925 distinct C\,{\sc iv} BAL 
troughs in the main sample. The $v_{\rm max}$ distribution of all BALs 
shows that this quantity is roughly distributed evenly between $-3000$ 
and  $-25000$~$\mathrm{km\,s^{-1}}$, while the corresponding 
$v_{\rm min}$ distribution  rises toward low velocities. The peak in the 
$v_{\rm min}$ distribution around $-3000$~$\mathrm{km\,s^{-1}}$ is 
artificially elevated due to the assigned lower velocity limit. The 
distribution histograms in Figures~10a and 10b show basic agreement 
with the similar distributions presented in G09 (see their Figure~8), 
although the $v_{\rm max}$ and $v_{\rm min}$ values in G09 give the 
velocities for all BALs together in a given spectrum.  
A K-S test for the $v_{\rm max}$ distributions of disappearing BAL troughs 
and all main-sample BAL troughs shows that these two distributions are 
not demonstrably inconsistent (K-S probability of 11.1\%), while the 
corresponding test using the $v_{\rm min}$ distributions gives a probability 
of 0.1\%. 
Based on this result and Figures~10a and 10b, it appears that BAL disappearance 
tends to occur for BAL troughs with relatively high values of $v_{\rm min}$; this 
result indicates that our requirement of trough detachment (see \S\ref{samples}) 
should not cause significant biases to our sample statistics.

Figures~10c and 10d show the distributions of the central velocity 
[$v_{\rm c}$ = \hbox{($v_{\rm max}$\,+\,$v_{\rm min}$)/2}] and trough width 
($\Delta v$ = $\mid v_{\rm max} - v_{\rm min}\mid$)  for disappearing and all 
925 distinct C\,{\sc iv} BAL troughs in the main sample. A K-S test yields a 
probability of 1.5\% for consistency between the two $v_{\rm c}$ distributions, 
indicating that disappearing BALs tend to have higher outflow central velocities 
than BALs in general.  A K-S test for the $\Delta v$ distributions of  
disappearing BAL troughs and all main-sample BAL troughs shows that these 
two distributions have a 6.9\% chance of consistency. These distributions 
suggest that disappearing BAL troughs tend to be narrower than the average 
BAL troughs.
  
\begin{figure}[h!]
\epsscale{1.15}
\plotone{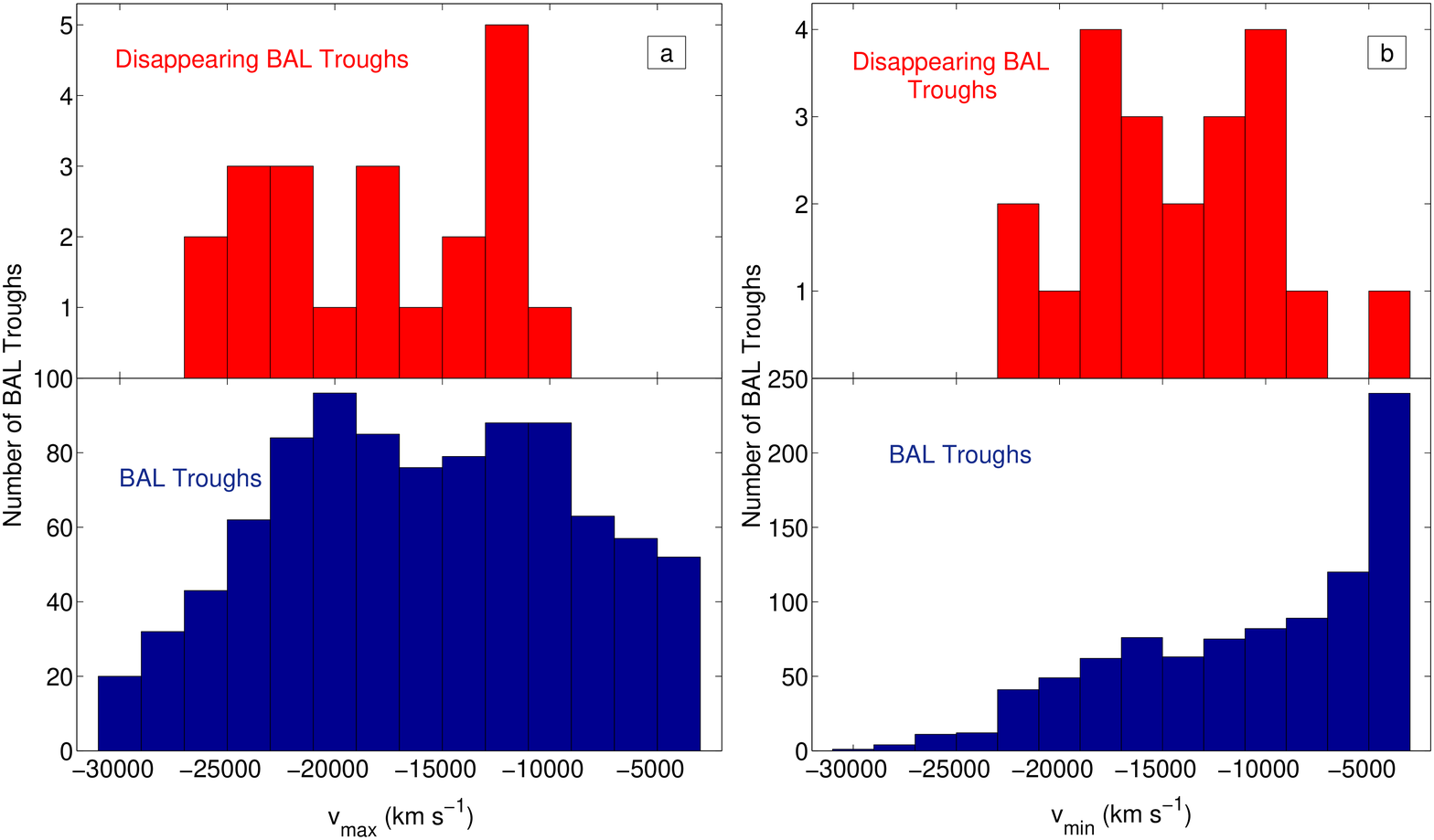}
\plotone{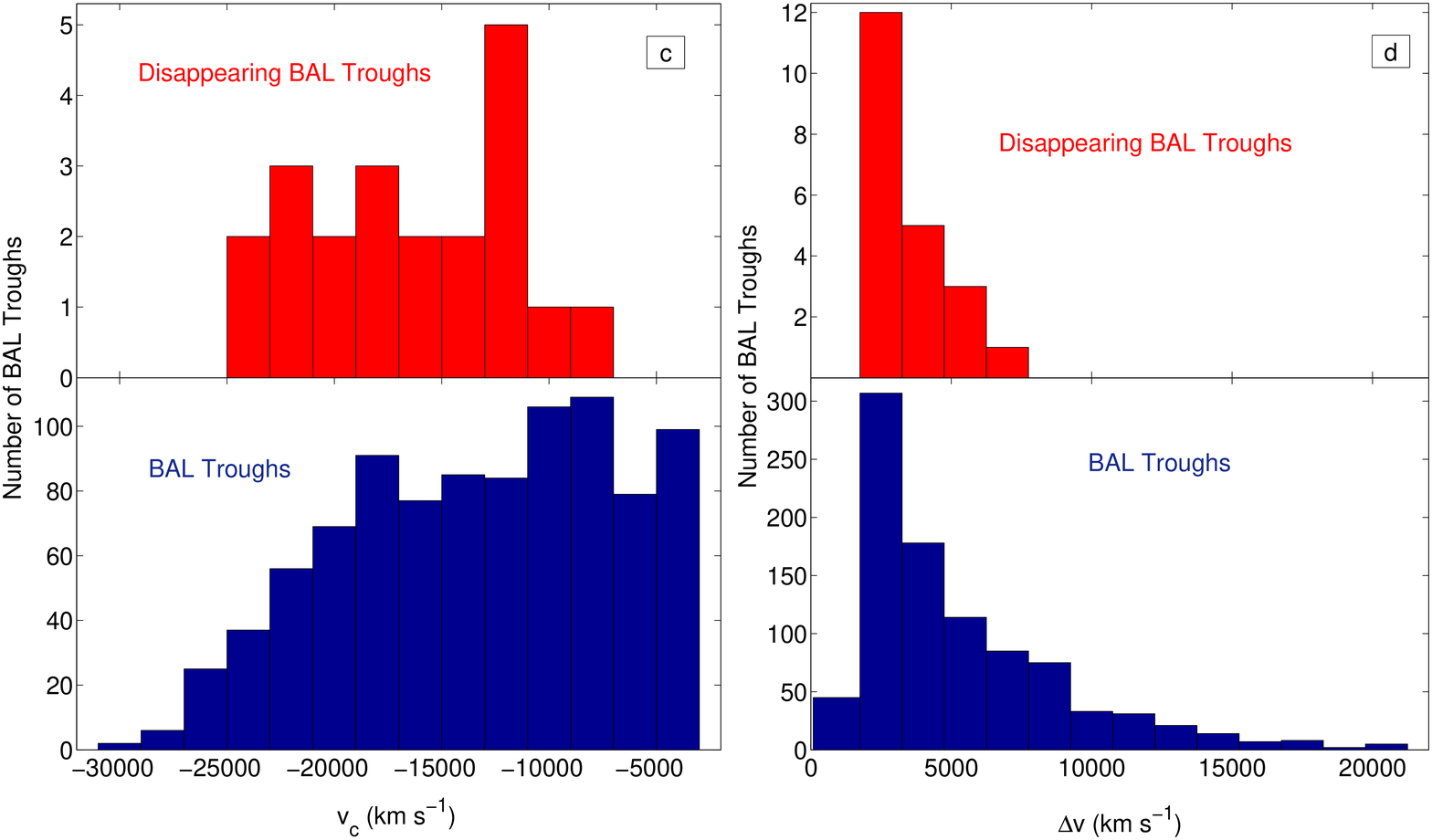}
\caption{$v_{\rm max}$ (a), $v_{\rm min}$ (b), $v_{\rm c}$ (c), and 
$\Delta v$  (d) distributions for disappearing C\,{\sc iv} BAL troughs 
(upper panels), and all 925 distinct C\,{\sc iv} BAL troughs in the main 
sample (lower panels).}
\label{fig10}
\end{figure}

According to our statistical tests above, C\,{\sc iv} BAL disappearance generally 
occurs for weaker troughs as well as higher velocity troughs. It is possible that 
these two results are related, or that one is simply the effect of the other. For 
example, in our main sample we find that weak BAL troughs can generally 
achieve higher velocities than strong BAL troughs (although there is a wide 
range of velocity observed at all trough strengths); as a result, the average
velocity for the population of weak BAL troughs is higher than that for strong 
BAL troughs. This basic result has also been noted by others 
(e.g., \citetalias{gibson09}; \citealp{cap11}). Thus, the tendency for BAL 
disappearance to occur for weaker troughs could also lead to disappearing 
troughs having higher velocities, on average, than for the whole trough 
population. A larger sample of disappearance events will be required to 
determine if BAL strength or BAL velocity, if either, is primarily connected to 
BAL disappearance.

\subsection{Connections Between Disappearing and Non-Disappearing Troughs} 
\label{connection}

Multiple troughs in the quasar spectra that have at least one disappearing 
BAL can be used to investigate connections between disappearing and 
non-disappearing troughs. Among the 19 quasars with disappearing troughs, 
we find nine that have at least one additional C\,{\sc iv} BAL trough that 
does not disappear (J093418.28+355508.3, 
J094806.58+045811.7, J104841.02+000042.8, 
 J112602.81+003418.2, J132216.24+052446.3, 
 J133152.19+051137.9, J134544.55+002810.7, 
 J142140.27$-$020239.0,  and J155119.14+304019.8). 
These nine quasars contain a total of 12 C\,{\sc iv} 
troughs that do not disappear and satisfy the criteria in \S\ref{samples}. 
We will hereafter refer to these as the ``additional non-disappearing C\,{\sc iv} 
troughs''. Figure~11 compares the EWs at two epochs, $t_1$ and  $t_2$, for 
distinct C\,{\sc iv} BAL troughs in the main sample and the additional 
non-disappearing C\,{\sc iv} troughs. The additional non-disappearing C\,{\sc iv} 
troughs almost always weaken (by up to 85\%), except for the least blueshifted 
trough of J133152.19+051137.9, which strengthens by 7\%. In 
J094806.58+045811.7, J104841.02+000042.8, J134544.55+002810.7, 
and J142140.27$-$020239.0, the additional non-disappearing C\,{\sc iv} 
trough transforms from a BAL to a mini-BAL as it weakens. The weakening of 
the additional non-disappearing C\,{\sc iv} troughs is statistically significant, 
given that the combinatorial probability to have 11 or more troughs out of 12 
weaken is only 0.3\%.  The fact that the additional non-disappearing troughs
usually weaken indicates that variability across multiple troughs is coordinated.

Figure~12 shows fractional EW variations for the 12 additional non-disappearing 
C\,{\sc iv} troughs versus central velocity offset from the disappearing trough. The 
error bars on EW fractional changes are calculated from the EW uncertainties at 
the two epochs (i.e., $t_1$ and $t_2$).  Notably, the coordinated variability across
multiple troughs persists even for velocity offsets as large as 
\hbox{10000--15000~km~s$^{-1}$}. This result is even more  remarkable given that 
BAL troughs often vary in discrete regions only a few thousand km~s$^{-1}$ wide 
\citep{gibson08}, and it requires further investigation. 

In Figure~12 nine of the 12 additional non-disappearing
C\,{\sc iv} troughs have smaller velocities than the disappearing C\,{\sc iv} troughs. 
This result implies that, for BAL quasars showing multiple troughs, the highest velocity 
trough is usually the one that disappears. This suggestive result is not clearly 
significant in the current sample, with a combinatorial probability of chance 
occurrence of 7\%, and it needs further investigation.

\begin{figure}[h!]
\epsscale{1.15}
\plotone{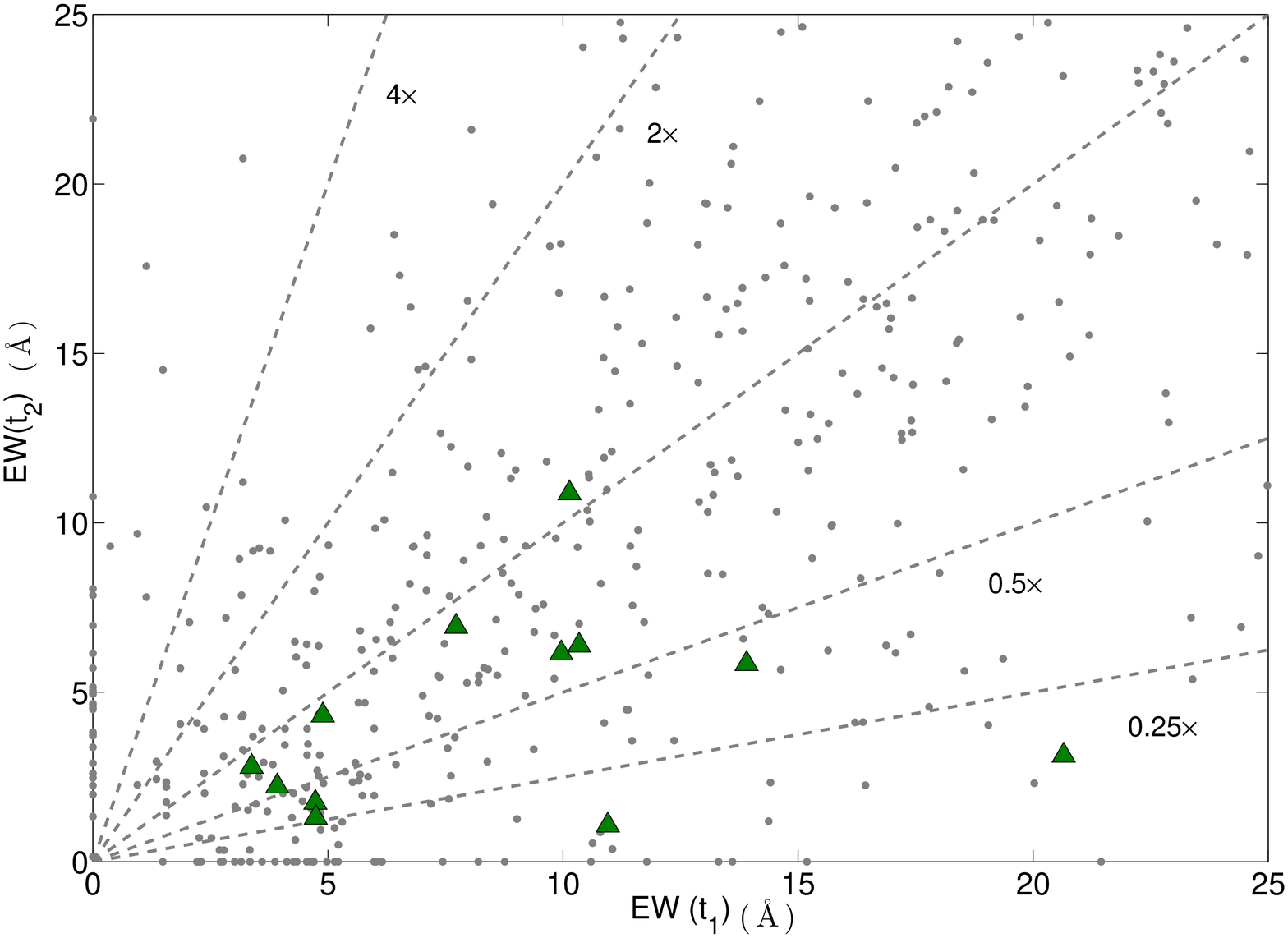}
\caption{The EWs at two epochs for all BAL troughs in the main sample (grey  
circles). The green triangles mark the other BAL troughs present in quasars that 
show one disappearing trough; note that in all cases but one the other BAL troughs 
weaken. Dashed lines indicate four times, two times, one-half of, and one-quarter 
of the first-epoch EWs, from top to bottom.}
\label{fig11}
\end{figure}

\begin{figure}[h!]
\epsscale{1.1}
\plotone{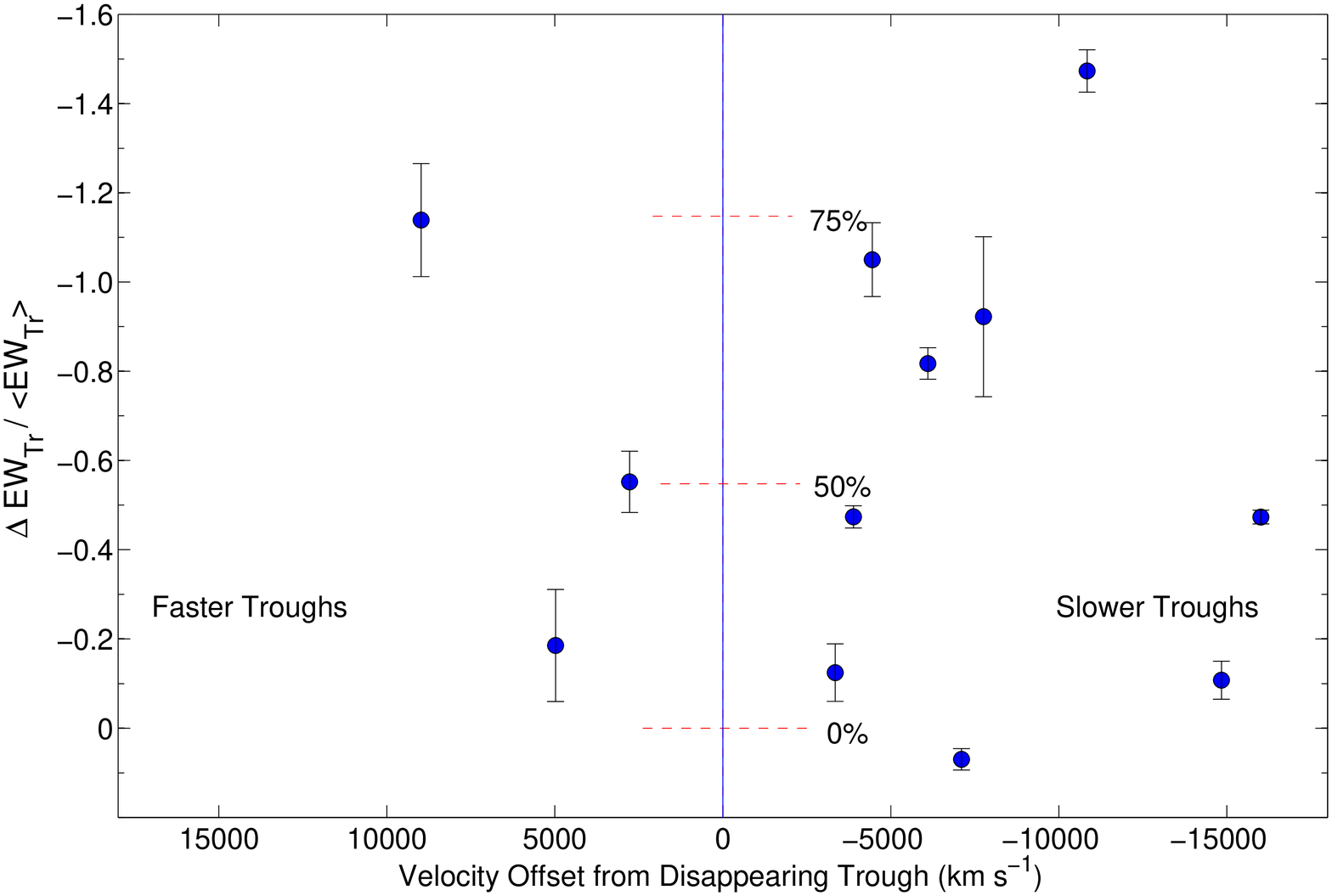}
\caption{Fractional changes in EW for 12 additional non-disappearing C\,{\sc iv} 
BAL troughs present in quasars that show one disappearing trough. The $x$-axis 
is the central velocity \hbox{[($v_{\rm max}$\,+\,$v_{\rm min}$)/2]} of each trough 
relative to that of the trough that disappeared. The horizontal dashed red lines 
indicate the levels for 50\%,  75\%, and 0\% decrease in trough strength. Note that 
even BAL troughs separated by 10000--15000\,$\mathrm{km\,s^{-1}}$ from the 
disappearing troughs weaken, showing coordinated variations over a large velocity 
range. }
\label{fig12}
\end{figure}

We also investigated Si\,{\sc iv} BAL troughs at velocities that correspond to one of 
the additional non-disappearing C\,{\sc iv} troughs. The additional Si\,{\sc iv} BAL 
troughs that are found in the spectra of three quasars show dramatic variations; the 
Si\,{\sc iv} BAL troughs in J132216.24+052446.3 and J142140.27$-$020239.0 
transform to mini-BALs (weakening in EW by 88\% and 87\%, respectively), and the 
trough in J155119.14+304019.8 disappears.

The results above demonstrate that outflow stream lines separated by large 
velocities apparently can either coordinate their variability or are being acted 
upon by some external agent that enforces coordination. The latter possibility 
appears more likely given understanding of BAL winds. One example of an 
external agent could be disk rotation. Such rotation might lead to coordinated
observed changes in absorption by several non-axisymmetric outflows that are 
loosely anchored to the accretion disk at different radii, provided there is some 
large-scale azimuthal asymmetry of the disk similarly affecting several stream 
lines. Another possible external agent could be the ``shielding gas'' that prevents 
BAL outflows from being overionized by highly energetic emission  generated 
close to the SMBH \citep[e.g.,][]{murray95,proga00}. This shielding gas is found 
to be variable in wind simulations \citep[e.g.,][]{sim10}, and the limited long-term 
X-ray variability observations show rare apparent examples of shielding-gas 
variations \citep[e.g.,][]{saez12}. Variations of the shielding gas could change 
the level  of
ionizing-continuum radiation reaching larger radii. In response to these changes 
in ionizing continuum, several absorption components at different velocities could 
rise and fall in ionization level together with some features disappearing. One 
challenge for this scenario is that BAL-trough variations generally do not  appear 
to be  correlated with variations of the observable continuum  (see \S\ref{intro}). 
However, since the observable continuum is typically that longward of 
Ly$\alpha$, it is perhaps possible that the ionizing continuum at shorter 
wavelengths varies differently in at least some cases. 

\section{Summary and Future Work} \label{summary}

We have used a systematically observed sample of 582 BAL quasars with 
925 distinct C\,{\sc iv} BAL troughs to provide the first statistically meaningful 
constraints upon BAL disappearance on multi-year timescales. Our main results 
are the following: 

\begin{enumerate}

\item
We have identified 21 cases of C\,{\sc iv} BAL disappearance in 19 quasars. 
On rest-frame timescales of \hbox{1.1--3.9~yr}, 
$f_{\rm disappear} = 2.3^{+0.6}_{-0.5}$\% of BAL troughs disappear and 
$f_{\rm quasar} = 3.3^{+0.9}_{-0.7}$\% of BAL quasars show a 
disappearing trough. 
If we consider only the pristine sample defined in \S \ref{selection}, then we find
11 cases of C\,{\sc iv} BAL disappearance in 11 quasars; the corresponding 
percentages are $f_{\rm disappear} = 1.2^{+0.4}_{-0.4}$\% and 
$f_{\rm quasar} = 1.9^{+0.8}_{-0.6}$\%.
See \S \ref{common}.

\item
The observed frequency of disappearing C\,{\sc iv} BAL troughs suggests  
an average trough rest-frame lifetime of 100--200 yr. See \S 
\ref{common}.

\item
Ten quasars showing C\,{\sc iv} BAL disappearance
have apparently transformed from BAL to non-BAL quasars;  these are the 
first reported examples of such transformations. The frequency
of BAL to non-BAL quasar transformation on timescales of 
\hbox{1.1--3.9~yr} is $1.7^{+0.7}_{-0.5} \%$.
See \S \ref{common}.

\item
The BAL quasars with disappearing troughs have representative
luminosities ($M_i$ values), SMBH mass estimates, and intrinsic 
reddening compared to our sample as a whole. See \S \ref{qsos}. 

\item 
As expected from the fact that most BAL quasars are radio quiet, 
most BAL quasars with disappearing troughs are radio quiet.
However, we do find one such quasar that is radio loud and another that is 
radio intermediate. Thus, BAL disappearance is a phenomenon of both 
radio-quiet  and radio-loud quasars. See \S \ref{qsos}.

\item
BAL disappearance appears to occur mainly for weak or 
moderate-strength absorption troughs but not the strongest 
ones; e.g., no troughs with EW~$>12$~\AA\ disappeared.
The BAL troughs that disappear are shallower than BAL troughs 
in general, although some fairly deep BAL troughs do
disappear. See \S \ref{ews}.

\item
Disappearing C\,{\sc iv} BAL troughs show higher outflow velocities than
BAL troughs in general, as indicated by their measured central
velocities and minimum velocities (though their measured maximum
velocities do not appear exceptional). There is also suggestive
evidence that disappearing BAL troughs tend to be narrower than
BAL troughs in general. This tendency for BAL disappearance to
occur for higher velocity troughs could be related to, or even
a secondary effect of, the fact that BAL disappearance appears to
occur mainly for weak or moderate-strength absorption troughs
(see point~5 above). See  \S \ref{ews}.

\item
When one BAL trough in a quasar spectrum disappears, the
other present troughs usually weaken (11 times out of 12 in our sample, 
corresponding to a significance level $>~99$\%).
The phenomenon occurs even for velocity offsets as large as 
\hbox{10000--15000 $\mathrm{km\,s^{-1}}$}. 
Variability across multiple troughs appears surprisingly
coordinated. Possible causes of such coordinated 
variations could be disk-wind rotation or variations of shielding gas that 
lead to variations of ionizing-continuum radiation.
These possible agents will need to be considered in future models
of quasar winds. See \S \ref{connection}.

\end{enumerate}

Given the results above, we can identify several promising observational
projects that should extend understanding of BAL disappearance. Further 
spectroscopy of the quasars that have shown BAL disappearance will allow
a search for reappearance of any of these BALs. Such reappearances
at the same measured velocities would not be expected if wind stream
lines have moved out of the line of sight owing to rotation of a non-axisymmetric 
outflow. However, BAL reappearance should be possible if the disappearance 
is a consequence of BAL weakening to strengths below our detection threshold. 
Systematic large-sample variability studies should let us assess the extent to 
which BAL disappearance is just the extension of normal BAL variability down 
to very small EWs. Further spectroscopy will also allow monitoring of the 
additional non-disappearing troughs. Furthermore, the planned absolute flux
calibration of the BOSS spectra  \citep[e.g.,][]{margala11} will allow a search 
for any systematic continuum-level changes associated with BAL disappearance.
The rate of BAL emergence events must balance that of BAL disappearance
events if the BAL quasar population is in a steady state, and thus
systematic large-scale studies of BAL emergence will be a critical
complement to those of disappearance. Finally, multiwavelength observations
of the quasars showing BAL disappearance are worthwhile. For example,
\hbox{X-ray} observations of objects that have transformed from BAL
to non-BAL quasars will be able to assess if the X-ray absorbing
shielding gas is still present along the line of sight.

The main-sample data set utilized in this study, along with the
still incoming BOSS observations, will be effective for a variety
of additional investigations of BAL variability. These include
studies of (1) absorption EW variability as a function of timescale
for different BAL transitions, (2) connections between BAL, emission-line,
and reddening variability, and (3) the effects of luminosity, redshift,
SMBH mass, Eddington fraction, and radio properties on BAL variability.

\acknowledgments

We gratefully acknowledge financial support from National Science Foundation 
grant AST-1108604 (NFA, WNB, DPS) and from NSERC (PBH). We  thank  D.~M. 
Capellupo  and F. Hamann for sharing their EW measurements for the \citet{cap11} 
sample in Figure 1. We thank  K. Dawson,  M. Eracleous,  D. Proga, and J. Wu for 
helpful discussions. We also thank the anonymous referee for useful feedback. 

Funding for SDSS-III has been provided by the Alfred P. Sloan Foundation, 
the Participating Institutions, the National Science Foundation, and the U.S. 
Department of Energy Office of Science. The SDSS-III web site is 
http://www.sdss3.org/.

SDSS-III is managed by the Astrophysical Research Consortium for the 
Participating Institutions of the SDSS-III Collaboration including the University 
of Arizona, the Brazilian Participation Group, Brookhaven National Laboratory, 
University of Cambridge, Carnegie Mellon University, University of Florida, the 
French Participation Group, the German Participation Group, Harvard University, 
the Instituto de Astrofisica de Canarias, the Michigan State/Notre Dame/JINA 
Participation Group, Johns Hopkins University, Lawrence Berkeley National 
Laboratory, Max Planck Institute for Astrophysics, Max Planck Institute for 
Extraterrestrial Physics, New Mexico State University, New York University, 
Ohio State University, Pennsylvania State University, University of Portsmouth, 
Princeton University, the Spanish Participation Group, University of Tokyo, 
University of Utah, Vanderbilt University, University of Virginia, University of 
Washington, and Yale University.

\clearpage

\begin{deluxetable}{crrccrc}
\tablecaption{Sample of Quasars Showing Disappearing BAL Troughs}
\tablewidth{0pt}
\tablehead{
\colhead{SDSS Name} & 
\colhead{Redshift\tablenotemark{a}}   &  
\colhead{$i$\tablenotemark{b}}   &  
\colhead{$M_i$\tablenotemark{c}}   &
\colhead{Plate-MJD-Fiber\tablenotemark{d}}   &  
\colhead{${\rm BI^\prime}$\tablenotemark{e}}   &  
\colhead{$N_{\rm Tr}$\tablenotemark{f}} \\
\colhead{} &  \colhead{$z$}  &  \colhead{(mag)}    &  \colhead{(mag)}    & 
\colhead{} 
& \colhead{$\mathrm{(km\,s^{-1})}$}   &  \colhead{} }
\startdata
J004022.40+005939.6\tablenotemark{$\ddagger$} & 2.565$\pm$0.0006 & 19.223$\pm$0.027 & $-$27.094 & 0690-52261-563 & 1668 & 1\\
 &  &  &  & 3587-55182-950 & 0 & 0\\
 &  &  &  & 3589-55186-558 & 0 & 0\\
 &  &  &  & 4222-55444-710 & 0 & 0\\
J074650.59+182028.7\tablenotemark{$\dagger$, $\ddagger$}  & 1.9163$\pm$0.0005 & 18.043$\pm$0.016 & $-$27.584 & 1582-52939-095 & 1241 & 1\\
 &  &  &  & 4492-55565-828 & 0 & 0\\
J081102.91+500724.2\tablenotemark{$\ddagger$} & 1.8422$\pm$0.0006 & 18.838$\pm$0.016 & $-$26.720 & 0440-51885-377 & 591 & 1\\
 &  &  &  & 0440-51912-395 & 665 & 1\\
 &  &  &  & 3699-55517-062 & 0 & 0\\
 &  &  &  & 4527-55590-028 & 0 & 0\\
J085904.59+042647.8\tablenotemark{$\dagger$, $\ddagger$} & 1.8104$\pm$0.0005 & 18.812$\pm$0.022 & $-$26.646 & 1192-52649-291 & 854 & 1\\
 &  &  &  & 3817-55277-538 & 0 & 0\\
 &  &  &  & 3814-55535-928 & 0 & 0\\
J093418.28+355508.3 & 2.4402$\pm$0.0007 & 18.902$\pm$0.016 & $-$27.400 & 1275-52996-096 & 1621 & 2\\
 &  &  &  & 4575-55590-498 & 1007 & 1\\
J093620.52+004649.2\tablenotemark{$\ddagger$} & 1.7213$\pm$0.0005 & 18.391$\pm$0.016 & $-$27.001 & 0476-52314-444 & 958 & 1\\
 &  &  &  & 3826-55563-542 & 0 & 0\\
J094806.58+045811.7 & 1.7371$\pm$0.0006 & 18.640$\pm$0.024 & $-$26.704 & 0994-52725-288 & 2146 & 3\\
 &  &  &  & 4798-55672-934 & 317 & 1\\
J104841.02+000042.8\tablenotemark{$\dagger$} & 2.0263$\pm$0.0006 & 18.720$\pm$0.016 & $-$26.970 & 0276-51909-310 & 2133 & 2\\
 &  &  &  & 3835-55570-398 & 0 & 0\\
J112602.81+003418.2 & 1.7928$\pm$0.0005 & 18.082$\pm$0.016 & $-$27.348 & 0281-51614-432 & 2421 & 2\\
 &  &  &  & 3839-55575-844 & 1024 & 1\\
J114546.22+032251.9\tablenotemark{$\ddagger$} & 2.0075$\pm$0.0007 & 19.058$\pm$0.023 & $-$26.721 & 0514-51994-458 & 389 & 1\\
 &  &  &  & 4766-55677-050 & 0 & 0\\
J132216.24+052446.3 & 2.0498$\pm$0.0006 & 18.384$\pm$0.019 & $-$27.438 & 0851-52376-622 & 2903 & 4\\
 &  &  &  & 4761-55633-794 & 1154 & 1\\
 &  &  &  & 4839-55703-442 & 1125 & 1\\
J133152.19+051137.9 & 1.7118$\pm$0.0005 & 18.182$\pm$0.019 & $-$27.150 & 0852-52375-626 & 3840 & 3\\
 &  &  &  & 4759-55649-756 & 3255 & 2\\
J133211.21+392825.9\tablenotemark{$\ddagger$} & 2.0520$\pm$0.0009 & 19.021$\pm$0.023 & $-$26.760 & 2005-53472-330 & 690 & 1\\
 &  &  &  & 4708-55704-412 & 0 & 0\\
J134544.55+002810.7 & 2.4680$\pm$0.0005 & 18.535$\pm$0.019 & $-$27.810 & 0300-51666-426 & 1712 & 1\\
 &  &  &  & 0300-51943-382 & 2875 & 1\\
 &  &  &  & 4043-55630-868 & 0 & 0\\
J142132.01+375230.3\tablenotemark{$\ddagger$} & 1.7791$\pm$0.0006 & 18.658$\pm$0.019 & $-$26.725 & 1380-53084-013 & 854 & 1\\
 &  &  &  & 4712-55738-030 & 0 & 0\\
J142140.27$-$020239.0 & 2.0878$\pm$0.0006 & 18.877$\pm$0.016 & $-$27.044 & 0917-52400-546 & 3950 & 3\\
 &  &  &  & 4032-55333-736 & 1002 & 1\\
J152149.78+010236.4\tablenotemark{$\ddagger$} & 2.2386$\pm$0.0004 & 18.558$\pm$0.018 & $-$27.602 & 0313-51673-339 & 807 & 1\\
 &  &  &  & 4011-55635-166 & 0 & 0\\
J152243.98+032719.8\tablenotemark{$\ddagger$} & 2.0002$\pm$0.0005 & 18.653$\pm$0.018 & $-$27.172 & 0592-52025-254 & 374 & 1\\
 &  &  &  & 4803-55734-442 & 0 & 0\\
J155119.14+304019.8 & 2.4104$\pm$0.0004 & 18.493$\pm$0.016 & $-$27.826 & 1580-53145-008 & 5176 & 2\\
 &  &  &  & 5011-55739-054 & 417 & 1\\
 &  &  &  & 5010-55748-492 & 382 & 1\\
 \enddata
\tablenotetext{a}{Redshifts are from \citet{hw10}, calculated from 
the cross correlation of the  Mg\,{\sc ii}, C\,{\sc iii}], and C\,{\sc iv} emission lines.}
\tablenotetext{b}{The $i$-band magnitude given in the SDSS DR5 quasar 
catalog (Schneider et~al. 2007).}
\tablenotetext{c}{Absolute $i$-band magnitude from Shen et~al. (2011). }
\tablenotetext{d}{Unique Plate-MJD-Fiber numbers  for each spectrum. BOSS 
spectra have MJD$\geq$55176 (see $\S$4 of \citealp{ross11}).}
\tablenotetext{e}{Balnicity index of each quasar in the given observation, 
summed over all troughs in the velocity range $-3000\,\geq\,v\,\geq\,-30000$\,$\mathrm
{km\,s^{-1}}$. None of the quasars in the main sample has a disappearing BAL 
trough beyond this velocity  range.}
\tablenotetext{f}{Number of BAL troughs in each spectrum.}
\tablenotetext{$\dagger$}{Blended NAL(s) with disappearing BAL troughs.}
\tablenotetext{$\ddagger$}{Quasars that transformed from BAL to non-BAL quasars.}
\end{deluxetable}

\clearpage
\begin{deluxetable}{ccrrrrrrr}
\tablewidth{0pt}
\tablecaption{Parameters of Disappearing BAL Troughs}
\tablehead{
\colhead{Name} & 
\colhead{MJD\tablenotemark{a}}   &  
\colhead{EW}   &  
\colhead{$v_{\rm max}$}   &  
\colhead{$v_{\rm min}$}   &  
\colhead{$f_{\rm deep}^{25}$\tablenotemark{b}}  &
\colhead{$\Delta t$\tablenotemark{c}} &
\colhead{$\log(P_{\chi^2}$)\tablenotemark{d}} 
\\
\colhead{SDSS} & 
\colhead{ }   &  
\colhead{($\mathrm{\AA}$)}   &  
\colhead{($\mathrm{km\,s^{-1}}$)}   &  
\colhead{($\mathrm{km\,s^{-1}}$)}   &
\colhead{ }   &  
\colhead{(days)}  &
\colhead{ }   }
\startdata
J004022.40+005939.6 & 52261 & 10.60$\pm$0.705 & $-$10067 & $-$4167 & 0.74 & 819.35 & $<-300$ \\
J074650.59+182028.7\tablenotemark{$\dagger$} & 52939 & 8.76$\pm$0.212 & $-$24994 & $-$18015 & 0.46 & 900.46 & $<-300$ \\
J081102.91+500724.2 & 51912 & 4.70$\pm$0.477 & $-$12405 & $-$9830 & 0.74 & 1268.38 & $-8.59$ \\
J085904.59+042647.8\tablenotemark{$\dagger$} & 52649 & 5.22$\pm$0.476 & $-$19315 & $-$16370 & 0.73 & 935.10 & $-11.34$ \\
J093418.28+355508.3 & 52996 & 3.60$\pm$0.272 & $-$25575 & $-$21959 & 0.31 & 754.03 & $-13.79$ \\
J093620.52+004649.2 & 52314 & 6.15$\pm$0.373 & $-$17603 & $-$13677 & 0.68 & 1193.91 & $<-300$ \\
J094806.58+045811.7 & 52725 & 5.98$\pm$0.549 & $-$21138 & $-$16797 & 0.55 & 1076.69 & $-15.96$ \\
J104841.02+000042.8\tablenotemark{$\dagger$} & 51909 & 2.25$\pm$0.185 & $-$12757 & $-$10654 & 0.33 & 1209.73 & $-11.51$ \\
J112602.81+003418.2 & 51614 & 4.27$\pm$0.148 & $-$26449 & $-$22919 & 0.54 & 1418.29 & $<-300$ \\
J114546.22+032251.9 & 51994 & 3.34$\pm$0.328 & $-$12771 & $-$9682 & 0.28 & 1224.61 & $-11.60$ \\
J132216.24+052446.3 & 52376 & 2.72$\pm$0.184 & $-$22727 & $-$20322 & 0.38 & 1067.94 & $<-300$ \\
                                          & 52376 & 3.02$\pm$0.218 & $-$18691 & $-$15564 & 0.26 & 1067.94 & $-9.29$ \\
                                          & 52376 & 2.30$\pm$0.176 & $-$13360 & $-$11213 & 0.33 & 1067.94 & $-9.90$ \\
J133152.19+051137.9 & 52375 & 3.16$\pm$0.219 & $-$12549 & $-$10440 & 0.56 & 1207.32 & $<-300$ \\
J133211.21+392825.9 & 53472 & 4.68$\pm$0.349 & $-$21377 & $-$17861 & 0.53 & 731.32 & $-12.41$ \\
J134544.55+002810.7 & 51943 & 7.60$\pm$0.182 & $-$11269 & $-$8096 & 0.92 & 1063.15 & $<-300$ \\
J142132.01+375230.3 & 53084 & 5.15$\pm$0.343 & $-$17005 & $-$14193 & 0.76 & 954.99 & $<-300$ \\
J142140.27$-$020239.0 & 52400 & 4.55$\pm$0.177 & $-$15968 & $-$12922 & 0.57 & 949.87 & $<-300$ \\
J152149.78+010236.4 & 51673 & 6.03$\pm$0.460 & $-$23658 & $-$18462 & 0.41 & 1223.37 & $<-300$ \\
J152243.98+032719.8 & 52025 & 2.77$\pm$0.203 & $-$14622 & $-$12300 & 0.34 & 1236.25 & $<-300$ \\
J155119.14+304019.8 & 53145 & 8.20$\pm$0.428 & $-$23588 & $-$17834 & 0.64 & 760.61 & $<-300$ \\
\enddata
\tablenotetext{a}{MJD of the observation used for BAL parameter measurements, 
which is taken to be the last SDSS observation that possesses the disappearing trough.}
\tablenotetext{b}{Fraction of BAL bins which lie at least  25\% under the continuum.}
\tablenotetext{c}{The rest-frame time interval between the last observation that 
possesses the disappearing trough, given in column 2, and the first 
observation that shows disappearance.}
\tablenotetext{d}{\,Logarithm of $\chi^{2}$ probability which gives the 
probability of  consistency between the SDSS and BOSS observations in the region 
limited by $v_{\rm max}$ and $v_{\rm min}$ (see \S\ref{selection}).}
\tablenotetext{$\dagger$}{Blended NAL(s) with disappearing BAL troughs.}
\end{deluxetable}

\begin{deluxetable}{lccccccccc}
\tablecaption{Observed Fractions and Lifetimes for Disappearing BAL Troughs}
\tablewidth{0pt}
\tablehead{
\colhead{ } & 
\colhead{$N_{\rm dt}$\tablenotemark{a}} & \colhead{${f_{\rm disapp}}$\tablenotemark{b}} & 
\colhead{$N_{\rm qdt}$\tablenotemark{c}}  & \colhead{${f_{\rm quasar}}$\tablenotemark{d} } & 
\colhead{$N_{\rm transform}$\tablenotemark{e}} & \colhead{$f_{\rm transform}$\tablenotemark{f}} & 
\colhead{${\bar{t}_{\rm trough}}$\tablenotemark{g}}  & \colhead{$\bar{t}_{\rm BAL}$\tablenotemark{h}}\\
\colhead{ } & \colhead{} & \colhead{(\%)} & \colhead{} & \colhead{(\%)} & 
\colhead{} &  \colhead{(\%)} & \colhead{(yr)} & \colhead{(yr)} }
\startdata
Standard Sample   &  21 & $2.3_{-0.5}^{+0.6}$ & 19 & $3.3_{-0.7}^{+0.9}$ & 
10 & $1.7_{-0.5}^{+0.7}$ & $109_{-22}^{+31}$ & $150_{-50}^{+60}$ \\[2ex]

Pristine Sample  &  11 & $1.2_{-0.4}^{+0.5}$ & 11 & $1.9_{-0.6}^{+0.8}$ & 
7  &  $1.2_{-0.4}^{+0.7}$ & $208_{-60}^{+105}$ & $208_{-80}^{+105}$ \\
\enddata  
\tablenotetext{a}{Number of disappearing C\,{\sc iv} BAL troughs}
\tablenotetext{b}{Fraction of disappearing C\,{\sc iv} BAL troughs}
\tablenotetext{c}{Number of quasars showing a disappearing C\,{\sc iv} BAL trough}
\tablenotetext{d}{Fraction of quasars showing a disappearing C\,{\sc iv} BAL trough}
\tablenotetext{e}{Number of quasars that transformed from BAL to non-BAL quasars}
\tablenotetext{f}{Fraction of quasars that transformed from BAL to non-BAL quasars}
\tablenotetext{g}{Average trough lifetime}
\tablenotetext{h}{\hbox{Lifetime} of the BAL phenomenon along our line of sight}
\end{deluxetable}

\end{document}